\documentclass[preprint1]{aastex63}
\usepackage{undertilde}

\newcommand{\kms}{km s$^{-1}$}

\newcommand{\h}{$^{\mbox{\scriptsize h}}$}
\newcommand{\m}{$^{\mbox{\scriptsize m}}$}

\usepackage{rotating}

\received{August 12, 2021}
\revised{September 28, 2021}
\accepted{October 18, 2021}
\submitjournal{ApJ}

\usepackage[version=3]{mhchem}
\usepackage{xcolor}
\usepackage{multirow}
\usepackage{amssymb}

\shorttitle{Mapping Orion KL at high spatial resolution}
\shortauthors{Wilkins et al.}

\begin{document}

\title{Mapping physical parameters in Orion KL at high spatial resolution}

\correspondingauthor{Olivia H. Wilkins}
\email{olivia.h.wilkins@outlook.com}

\author[0000-0001-7794-7639]{Olivia H. Wilkins}
\affiliation{Division of Chemistry and Chemical Engineering, California Institute of Technology, Pasadena, CA 91125 USA}

\author[0000-0002-3191-5401]{P. Brandon Carroll}
\affiliation{Center for Astrophysics $\mid$ Harvard \& Smithsonian, Cambridge, MA 02138 USA}

\author[0000-0003-0787-1610]{Geoffrey A. Blake}
\affiliation{Division of Chemistry and Chemical Engineering, California Institute of Technology, Pasadena, CA 91125 USA}
\affiliation{Division of Geological and Planetary Sciences, California Institute of Technology, Pasadena, CA 91125 USA}

\begin{abstract}


The Orion Kleinmann-Low nebula (Orion KL) is notoriously complex and exhibits a range of physical and chemical components. We conducted high angular resolution (sub-arcsecond) observations of \ce{^13CH3OH} $\nu=0$ (${\sim}0\farcs3$ and ${\sim}0\farcs7$) and \ce{CH3CN} $\nu_8=1$ (${\sim}0\farcs2$ and ${\sim}0\farcs9$) line emission with the Atacama Large Millimeter/submillimeter Array (ALMA) to investigate Orion KL's structure on {small} ~spatial scales (${\le}350$ au). Gas kinematics, excitation temperatures, and column densities were derived from the molecular emission via a pixel-by-pixel spectral line fitting of the image cubes, enabling us to examine the {small-scale} ~variation of these parameters. Sub-regions of the Hot Core have a higher excitation temperature in a $0\farcs2$ beam than a $0\farcs9$ beam, indicative of possible internal sources of heating. Furthermore, the velocity field includes a bipolar ${\sim}7{-}8$ \kms ~feature with a southeast-northwest orientation against the surrounding ${\sim}4{-}5$ \kms ~velocity field, which may be due to an outflow. 
We also find evidence of a possible source of internal heating toward the Northwest Clump, {since the excitation temperature there is higher in a smaller beam versus a larger beam.} 
Finally, the region southwest of the Hot Core (Hot Core-SW) presents itself as a particularly heterogeneous region bridging the Hot Core and Compact Ridge. Additional studies to identify the (hidden) sources of luminosity and heating within Orion KL are necessary to better understand the nebula and its chemistry.

\end{abstract}




\section{Introduction} \label{sec:intro}

\begin{figure}
    \centering
    \includegraphics[width=0.45\textwidth]{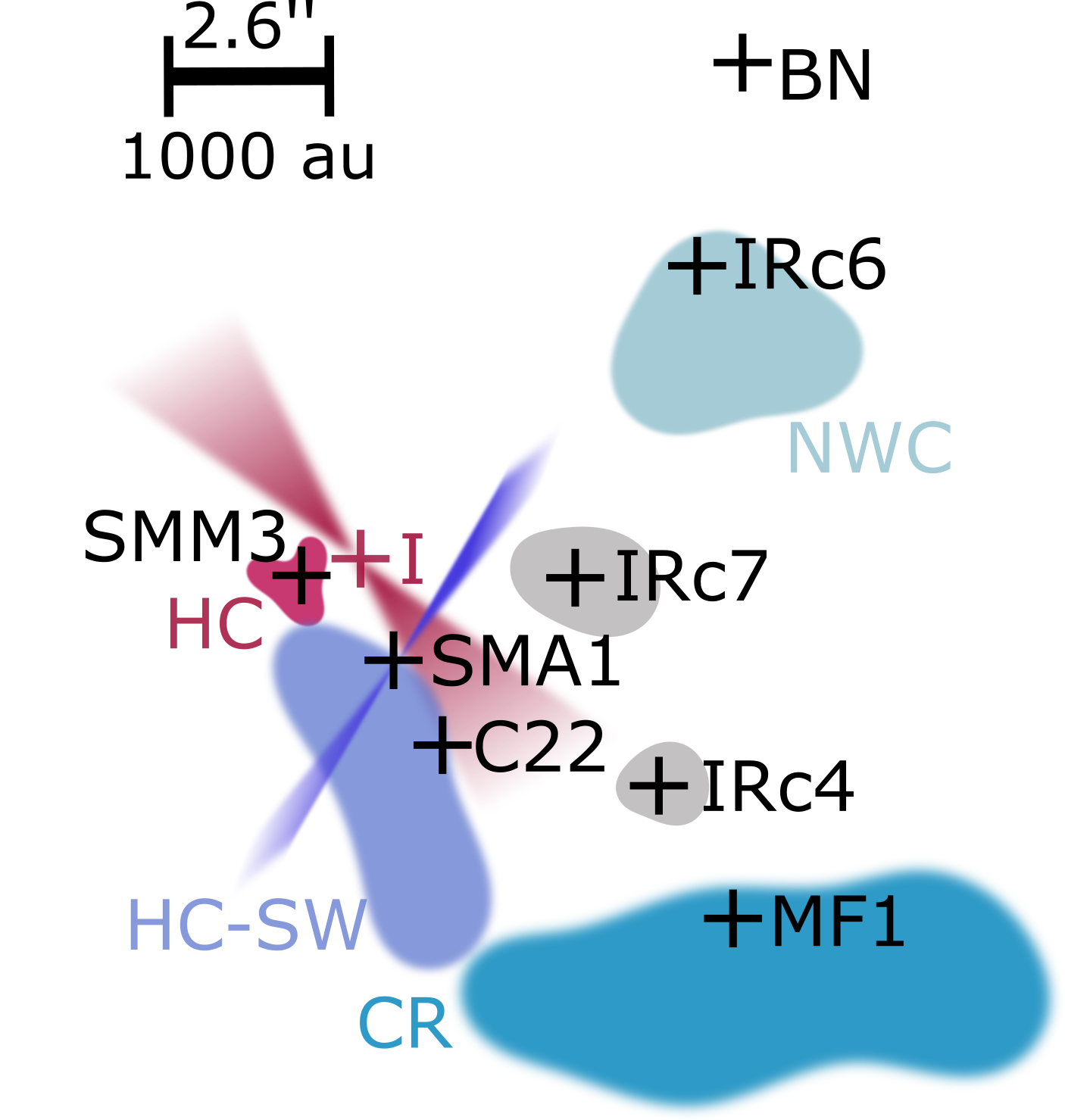}
    \caption{Illustration of Orion KL's morphology, as seen in the plane of the sky. Regions of interest include Source I (I), the Hot Core (HC), the Hot Core-Southwest (HC-SW), the Compact Ridge (CR) and the Northwest Clump (NWC). The crosses indicate specific emission regions that are discussed in the text. The scale bar in the top left corresponds to a linear scale of $\sim$1000 au (2.6\arcsec) at Orion KL's distance of 388 pc.}
    \label{fig:morphology}
\end{figure}

{Molecules are useful tools for characterizing the physical structure of dense interstellar environments} \citep[e.g., ][]{Herbst2009, Ginsburg2017,Moscadelli2018,Gieser2019,Law2021}. {As has been illustrated repeatedly with the Orion Kleinmann-Low nebula (Orion KL, $d\sim 388$ pc)---the closest region of high-mass star formation to the Earth \citep{Kounkel2017}---, different types of molecules can be used to trace different types of environments and conditions in interstellar material}. {In Orion KL,} the complex interplay of dense molecular gas and star formation is traced by several components, such as the Hot Core, the Compact Ridge, the Extended Ridge, and the Plateau regions, each of which has varying chemical and physical properties \citep[e.g., ][]{Blake1987,Friedel2011,Feng2015,Tercero2018,Luo2019,Cortes2021}. {Figure~\ref{fig:morphology} shows a schematic representation of Orion KL illustrating regions discussed in this work.} The variety of {chemical} and physical properties, as well as the {readily detectable emission lines from} isotopologues \citep[e.g., ][]{Neill2013}, observed toward Orion KL make it an excellent laboratory for studying the formation and subsequent chemistry of molecules in high-mass star-forming regions. 

Generally, {the two dominant sub-regions of} Orion KL that harbor complex chemistry are the {so-called ``Hot Core''} which contains denser and warmer {gas} ($n_{\mbox{\scriptsize \ce{H2}}}\sim 10^7$ cm$^{-3}${, $T_{kin}\sim 200$ K}) and the {``Compact Ridge''}  that lies to the southwest of the Hot Core, and which is both cooler and less dense ($n_{\mbox{\scriptsize\ce{H2}}} \utilde{<} 10^6$ cm$^{-3}${, $T_{kin}\sim 100{-}150$ K}) \citep{Blake1987,Genzel1989}. The regional variation in physical conditions has also resulted in large-scale variation in the chemistry, with the Hot Core and Compact Ridge being the sites of {prominent emission from} nitrogen-bearing and oxygen-bearing species, respectively {\citep[e.g.][]{Blake1987,Friedel2011}}.

The variation in both the physical conditions and chemistry of Orion KL has been well-documented across various spatial scales, e.g. $\sim$20\arcsec$-$43\arcsec ~with \textit{Herschel/HIFI} by \citet{Wang2011}; $\sim$0\farcs5$-$5\farcs0 with CARMA by \citet{Friedel2011}; and $\sim$2\farcs6$-$3\farcs7 with combined SMA and IRAM 30m data by \citet{Feng2015}. {There has also been high-{angular-}resolution {millimeter} continuum {imaging of} the nebula, e.g. by \citet{Hirota2015} at $\sim$0\farcs5} with ALMA. \citet{Ginsburg2018} and \citet{Wright2020} imaged continuum and line emission specifically toward the disk and outflow of Source I alone down to 0\farcs03.

Observations of molecular line emission have shed light on both the velocity structure and sources of heating within Orion KL. Ammonia (\ce{NH3}) and methyl formate (\ce{HCOOCH3}), for example, have been used to trace the velocity fields throughout the nebula at sub-arcsecond angular scales \citep[e.g., ][]{Goddi2011,Pagani2017}. \ce{CH3OH}, \ce{HCOOCH3}, thioformaldehyde (\ce{H2CS}), and cyanoacetylene (\ce{HC3N}) are included in the suite of molecules used to probe whether Orion KL's different components are internally or externally heated, a question that remains the subject of debate toward the Hot Core specifically \citep{deVicente2002,Goddi2011,Zapata2011,Crockett2014,OrozcoAguilera2017,Peng2017,Li2020}.

Even with these studies, many questions remain about the nature of Orion KL. Investigations of the nebula's three-dimensional (3-D) structure, for example, have mostly looked at the so-called high-energy ``fingers" and ``bullets" emanating on arcminute scales from an explosion that took place in the nebula about 500 years ago \citep[][]{Nissen2007,Bally2015,Youngblood2016,Bally2017}. However, a more {small-scale} view of the 3-D structure of the dense molecular gas in Orion KL is largely absent in the literature, despite continuum and molecular line emission imaging showing that both the physical and chemical complexity of Orion KL are not simply regional but are manifest on more localized scales as well. {That is, while the physical and chemical aspects of Orion KL can be divided into large areas such as the dense, nitrogen-rich Hot Core and the less dense, oxygen-rich Compact Ridge, there is growing evidence suggesting that this heterogeneity extends to within these subregions of the nebula as well} \citep[e.g., ][]{Friedel2011,Crockett2014}. 

Thus far, the results of high-angular-resolution imaging of Orion KL have generally focused on distinct regions or line emission peaks. Such observations at various spatial scales are imperative to understanding the physical and chemical conditions of star-forming regions, and they reveal that there is still much to uncover in Orion KL. 

Here we present a more localized ($\sim$0\farcs9-0\farcs2) view of Orion KL using a combination of ground-state \ce{^13CH3OH} and vibrationally-excited \ce{CH3CN} line emission with the Atacama Large Millimeter/submillimeter Array (ALMA). {These compounds are both complex (i.e., have at least six atoms) and (near) symmetric top molecules, which means they have spectra that are both abundant in lines and relatively simple, such that deriving their abundances and temperatures is rather straightforward. We specifically target methanol to trace colder, less dense gas; the carbon-13 methanol isotopologue was specifically targeted because it serves as an optically thin proxy for the primary carbon-12 isotopologue, which is optically thick toward Orion KL. Similarly, \ce{CH3CN} $\nu_8=1$ was selected to trace hotter and denser gas, to probe high energy emission excited by massive embedded protostars, and because ground-state \ce{CH3CN} is optically thick toward the Orion KL Hot Core.}

The observations presented here complement previous studies of Orion KL to further elucidate the small-scale physical structure of the nebula. {The angular resolutions (from 0\farcs9 down to 0\farcs2) were deliberately chosen to enable observations of} \edit1{small-scale, sub-region variations in the chemical and physical conditions} {at the expense of resolving out extended emission that is already widespread in the existing literature.} Combining both {molecular} tracers in a single analysis is needed to characterize the wide range of environments in Orion KL, especially in light of the longstanding observations that nitrile and oxygen-bearing organic emission lines are kinematically and spatially distinct, even at 10-30$''$ spatial resolution \citep[e.g.][]{Blake1987,Crockett2014}. We find that our observations agree overall with previous studies toward Orion KL; however, we resolve spatial structure at scales on the order of protoplanetary disks, which lie within radii typically associated with molecular abundance enhancements caused by heating from embedded protostars \citep{Boonman2001,Schoeier2002}. Furthermore, our results provide new insights to the kinematics and thermal profile of the nebula. 

{The ALMA observations and an overview of the data analysis are described in Sections~\ref{sec:obs} and~\ref{sec:methods}, respectively. The chemical distribution, including column density measurements, is discussed in Section~\ref{sec:chemdist}. A discussion of the physical and chemical structure in Orion KL, including line width and velocity field maps, is presented in Section~\ref{sec:structure}. Section~\ref{sec:thermalstructure} provides a discussion of derived temperature maps and possible sources of heating. The results of this work are summarized in Section~\ref{sec:summary}.} \\

\section{Observations}\label{sec:obs}

Observations of Orion KL were taken in ALMA Cycles 4 and 5 with a pointing center of $\alpha_{\mbox{\scriptsize J2000}} = 05$\h 35\m 14\fs50, $\delta_{\mbox{\scriptsize J2000}} = -05\arcdeg 22\arcmin 30\farcs9$. {All observations were carried out \edit1{with} the 12-meter array.}

The Cycle 4 observations (project code \#2016.1.01019, PI: Carroll) include the \ce{CH3CN} $\nu_8=1$ {lines} used in these analyses. {These observations comprised two epochs, each with one execution block and and one spectral window covering a frequency range of 146.83-147.77 GHz in Band 4 at a spectral resolution of 488.28 kHz ($\sim$1.0 km s$^{-1}$). The Cycle 4 data taken on 2 October 2016 used 44 antennas with projected baselines between 18.6 m and 3.2 km (9.3k$\lambda$ and 1600k$\lambda$) and a primary beam (field of view) of 39.8\arcsec. The on-source integration time was 455 s. The precipitable water vapor was 2.1 mm, and typical system temperatures were around 50-100 K. The observations taken on 28 November 2016 used 47 antennas with projected baselines between 15.1 m and 704.1 m (7.6k$\lambda$ and 350k$\lambda$) and a primary beam of 39.8\arcsec. The on-source integration time was 151 s. The precipitable water vapor was 1.9 mm, and typical system temperatures were around 50-100 K.}

The Cycle 5 observations (project code \#2017.1.01149, PI: Wilkins) include the \ce{^13CH3OH} $\nu=0$ lines used in these analyses. {These observations comprised one execution block and 10 spectral windows, three of which contained the targeted \ce{^13CH3OH} lines. These three windows covered frequency ranges of 155.13-155.37 GHz, 155.57-155.80 GHz, and 156.06-156.29 GHz in Band 4 at spectral resolutions of 244 kHz ($\sim$0.5 km s$^{-1}$). These observations were carried out on 14 December 2017 and used 49 antennas with projected baselines between 15.1 m and 3.3 km (7.6k$\lambda$ and 1650k$\lambda$) and a primary beam of 39.1\arcsec. The on-source integration time was 2062 s. The precipitable water vapor was 3.7 mm, and typical system temperatures were around 75-125 K.} 

The spectra for each molecular probe analyzed here were confined to a single correlator sub-band; as such, the uncertainties for quantities derived from for these molecular probes are dominated by the thermal and phase noise and are mostly unaffected by systemic calibration uncertainties. Calibration was completed using standard CASA calibration pipeline scripts using CASA 4.7.0 for Cycle 4 data and CASA 5.1.1-5 for Cycle 5 data. \edit1{For both sets of observations, the} source J0423$-$0120 was used as a calibrator for amplitude, atmosphere, bandpass, pointing, and WVR (Water Vapor Radiometer) variations, while J0541$-$0211 was used as a phase and WVR calibrator. 

The image cubes presented here were created with continuum emission estimated from line-free channels subtracted in the UV-plane using the \texttt{uvcontsub} task in CASA. 

\subsection{Cycle 4 Data Reduction}

The Cycle 4 data were processed {in CASA 4.7.0} using the{ Cycle 4 ALMA data processing pipeline,\footnote{\edit1{\citep[Pipeline-Cycle4-R2-B, r38377, ][]{pipelinecyc4}}} which used the \texttt{tclean}} algorithm with Briggs weighting and {a pipeline-generated mask} \edit1{made by identifying pixels likely to contain emission lines}. {A robust parameter of 0.5 was used} for deconvolution. The resulting continuum and line images are primary-beam-corrected and have a noise-level of{ $\sigma_{\mbox{\scriptsize RMS}}\sim 1.2$ mJy beam$^{-1}$.}
Comprising two epochs, that yield images of \ce{CH3CN} $\nu_8=1$ at two angular resolutions, the Cycle 4 data taken on 2016 October 2 have a 0\farcs23$\times$0\farcs21 synthesized beam (hereafter identified using the shorthand 0\farcs2), which corresponds to linear scales of $\sim$90 au in Orion KL{, with a position angle of PA $=$ -8$^\circ$}. Images from observations on 2016 November 28 have a synthesized beam of 1\farcs13$\times$0\farcs72 (hereafter 0\farcs9), which corresponds to linear scales of $\sim$350 au in Orion KL{, with PA $=$ -71$^\circ$}. 

{The targeted \ce{CH3CN} $\nu_8=1$ lines in these data were also detected in Cycle 2 observations (project \#2013.1.01034, PI: Crockett, \citet{Carroll2017thesis}) toward Orion KL but at an angular resolution of $\sim$2\farcs0 ($\sim$800 au). Comparing the integrated intensities for the \ce{CH3CN} lines listed in Table~\ref{tab:lines}, we find that we recover $\sim$28\% and $\sim$16\% of the flux integrated over the whole Orion KL region in the 0\farcs2 and 0\farcs7 images, respectively, compared to what is recovered in the 2\farcs0 observations.}

\subsection{Cycle 5 Data Reduction}\label{sec:obscyc5}

The Cycle 5 data were similarly processed {in CASA 5.1.1-5} using the \texttt{tclean} algorithm {(interactive)} with Briggs weighting{ and the `auto-multithresh' masking algorithm \citep[][]{Kepley2020}}. A robust parameter of 1.5 (i.e., semi-natural weighting) was used for deconvolution to overcome significant artifacts {that obstructed analysis of the images when using a robust parameter of 0.5. One possible explanation for these artifacts is that the \ce{^13CH3OH} emission is extended such that it is difficult to image at longer baselines, thus requiring a robust parameter that gave higher weights to shorter baselines.} The images are primary-beam-corrected and have a noise-level of $\sigma_{\mbox{\scriptsize RMS}}\sim 1.0$ mJy beam$^{-1}$.

There was only one set of observations taken, on 14 December 2017, toward Orion KL during the Cycle 5 program. These images have a synthesized beam of 0\farcs34$\times$0\farcs20 (hereafter 0\farcs3), which corresponds to linear scales of $\sim$110 au in Orion KL{, with PA $=$ -56$^\circ$}. To compare the \ce{^13CH3OH} emission on multiple spatial scales, we re-imaged these data {splitting the measurement set to include} only baselines of $\le$500 m, resulting in a synthesized beam of 0\farcs74$\times$0\farcs63 (hereafter $0\farcs7$), which corresponds to linear scales of $\sim$270 au in Orion KL{, with PA $=$ -72$^\circ$}. The image cube generation was otherwise done in the same fashion as that used for the full \ce{^13CH3OH} data set. 

The $J=5$ transition of \ce{^13CH3OH} at 156.3 GHz in these data was also detected in {the aforementioned} Cycle 2 observations toward Orion KL but at $\sim$2\farcs0 angular resolution ($\sim$800 au). Comparing the integrated intensities for this line, we find that we recover $\sim$9\% and $\sim$14\% of the \ce{^13CH3OH} flux {integrated over the whole Orion KL region} in the 0\farcs3 and 0\farcs7 images, respectively, compared to {what is recovered in} the 2\farcs0 observations. {Moreover, we estimate that we recover about 4\% and 6\% of the total \ce{^13CH3OH} $J=5$ flux based on rough estimates of the flux derived from abundances and rotation temperature measurements made using \textit{Herschel} \citep{Crockett2014}. We attribute this to extended \edit1{emission} being resolved out in our observations, which have maximum recoverable scales equivalent to about 930-2400 au in Orion KL.} In this work, however, we focus on probing the {small-scale} structure rather than large-scale features of the nebula. Thus we emphasize that these observations complement the existing lower-angular-resolution studies of Orion KL and that observations at multiple spatial scales are imperative for a holistic picture of star-forming regions such as Orion KL. 

\section{Methods}\label{sec:methods}

As noted in Section 1, different molecular tracers can be used as probes of different conditions in star-forming regions. The spectra from the ten spectral windows in the ALMA Cycle 5 observations, which are shown in Figure~\ref{fig:spectrumfull}, illustrate this, with distinct profiles observed in the Hot Core (red) and the Compact Ridge (blue). {While some transitions emit with similar intensities across the nebula, some lines show up in either the Hot Core or the Compact Ridge, but not both, while others have lines in both regions but at substantially different intensities.} Using three transitions of \ce{^13CH3OH} and eight transitions of \ce{CH3CN} $\nu_8=1$, which we list in Table~\ref{tab:lines}, we resolve detailed profiles of velocity, temperature, and line widths of these tracers toward Orion KL. {As discussed in Section~\ref{sec:obs}, these lines were chosen because they were observed within a single correlator set-up (across three spectral windows for \ce{^13CH3OH} and in one spectral window for \ce{CH3CN} $\nu_8=1$), hence minimizing the uncertainties for the derived parameters. While there were additional lines present in the data, the selected lines represent a subset of consistently strong and unblended emission lines across the nebula.} 

\begin{figure*}[b!]
\begin{center}
\includegraphics[width=\textwidth]{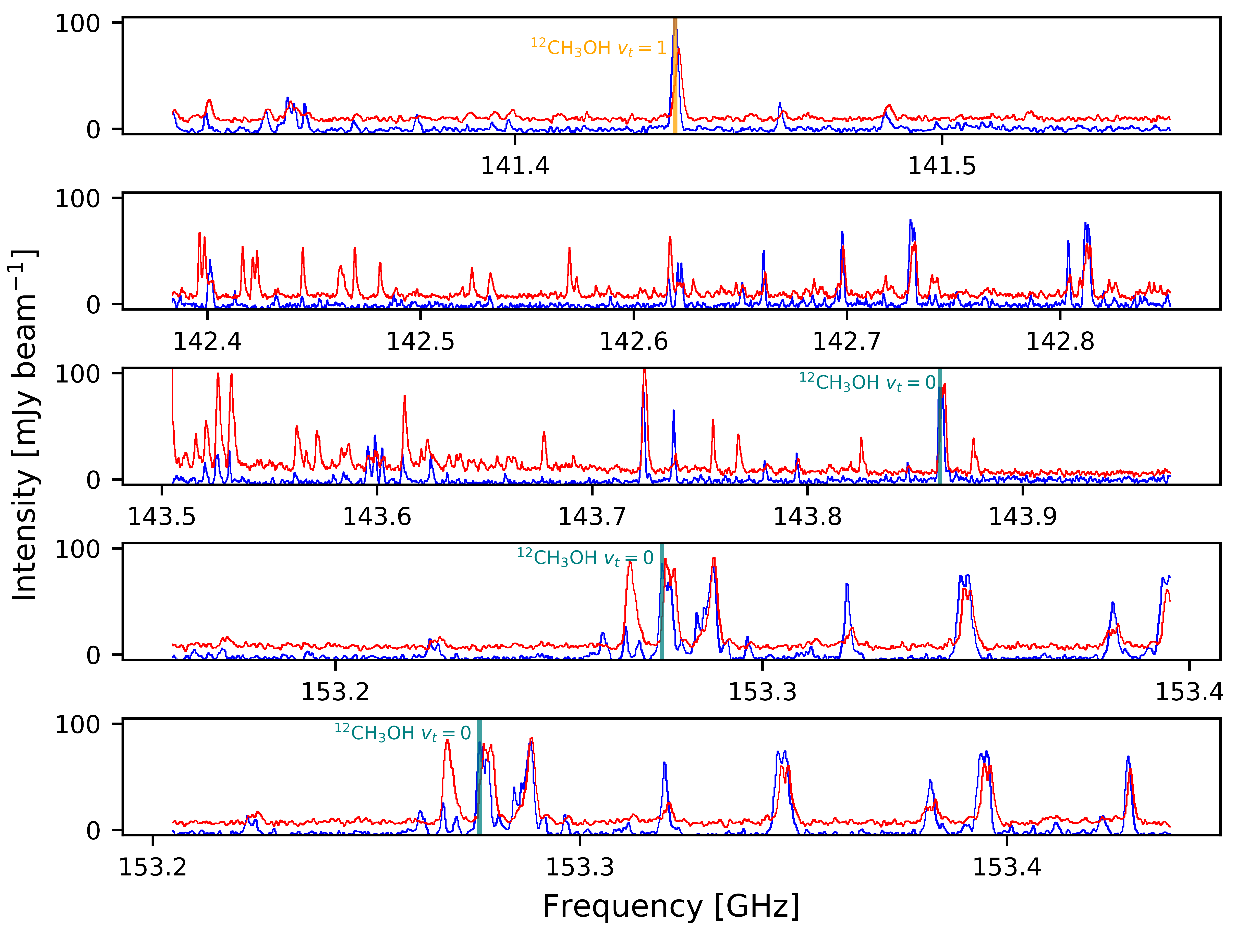}
\caption{Spectra from ten ALMA Cycle 5 Band 4 spectral windows toward Orion KL, displaying the plethora of molecular line emission toward the Compact Ridge (blue) and the Hot Core (red, offset by 10 mJy beam$^{-1}$ for better visibility). Solid lines show transitions of \ce{^13CH3OH} (black), \ce{^12CH3OH} $v_t=0$ (teal), and  \ce{^12CH3OH} $v_t=1$ (orange).{ All tick marks are separated by 0.1 GHz. \textit{Continued on next page.}}}
\label{fig:spectrumfull}
\end{center}
\end{figure*}

\begin{figure*}[t!]
\figurenum{2}
\begin{center}
\includegraphics[width=\textwidth]{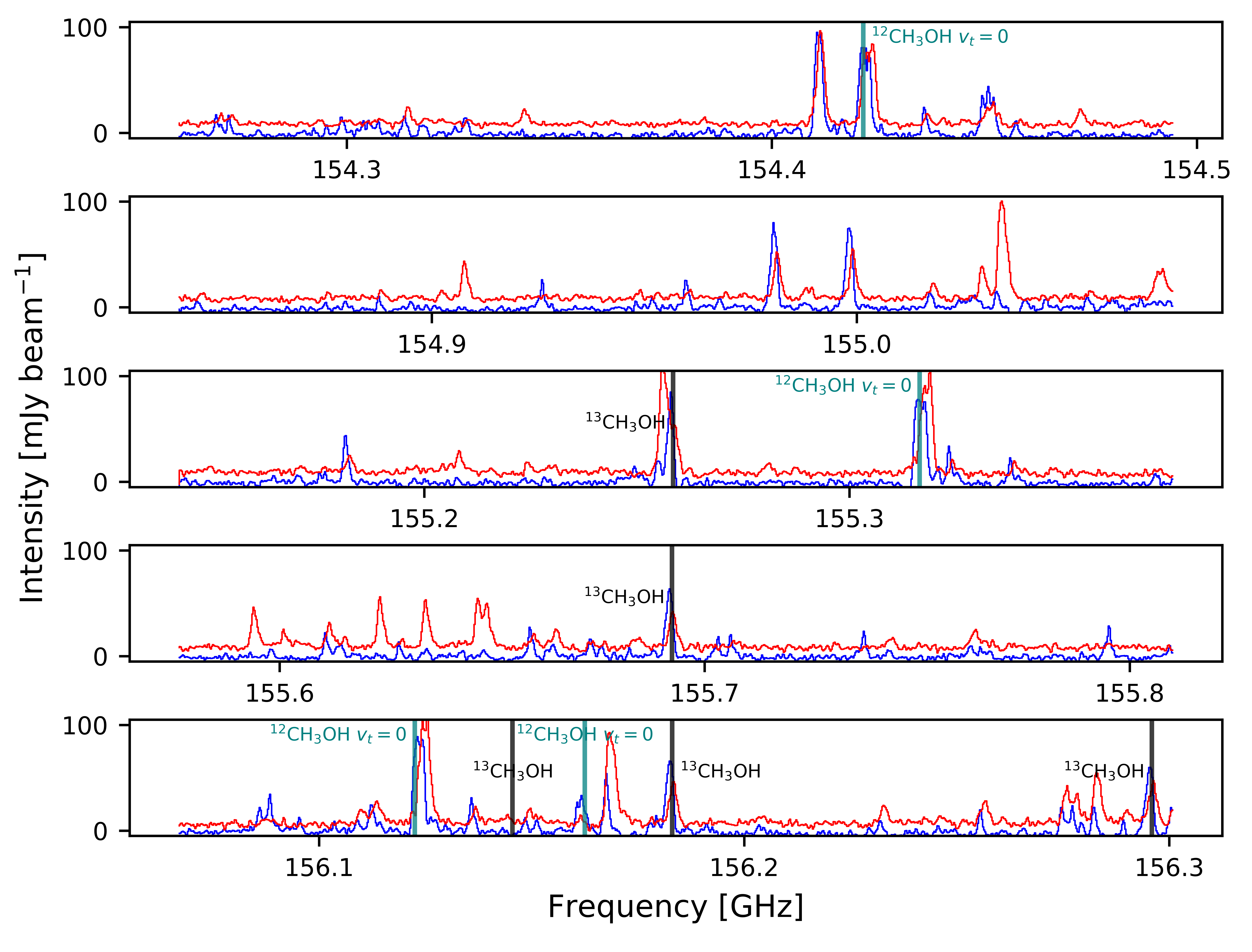}
\caption{\textit{Continued from previous page.}}
\end{center}
\end{figure*}

Line width, velocity, excitation temperature, and column density of \ce{^13CH3OH} $\nu=0$ and \ce{CH3CN} $\nu_8=1$ as a function of position were derived by a pixel-by-pixel fit of the data.\footnote{Python script available at \href{https://github.com/oliviaharperwilkins/LTE-fit}{https://github.com/oliviaharperwilkins/LTE-fit}.} Because all lines used in the fit for each molecule were observed simultaneously, the uncertainties in excitation temperature, which are derived from relative fluxes, should be dominated by thermal noise rather than by multiple sources of calibration uncertainty. For the column density estimates, the relative differences across the images should again be dominated by thermal (and phase) noise, while the overall calibration error budget does impact the total column densities derived {but by $\le$5\% in Band 4 \citep[][]{pg}.}

The pixel-by-pixel fit of the data was achieved by extracting spectra in a single synthesized beam centered on each pixel, in succession. To avoid erroneous fits, only lines with {peak amplitudes} of $\ge$3$\sigma_{\mbox{\scriptsize RMS}}$ were considered. Line parameters---namely line width, velocity shift, excitation temperature, and column density---were determined simultaneously by least-squares fitting with \texttt{LMFIT}\footnote{\texttt{LMFIT} {\citep[][]{lmfit}} is a non-linear least-squares minimization and curve-fitting algorithm for Python that can be accessed at \href{https://doi.org/10.5281/zenodo.598352}{https://doi.org/10.5281/zenodo.598352}.} {(version 0.9.15)} to model the spectra assuming {optically thin lines at} local thermodynamic equilibrium (LTE){\footnote{For the rationale behind the assumption of optically thin lines at LTE, see Appendix~\ref{sec:LTE}.}} using Equation~\ref{eq:Ntot} \citep[adapted from][]{Remijan2003}:
\begin{equation}\label{eq:Ntot}
N_{tot} = 2.04\times 10^{20} C_{\tau} \frac{1}{\Omega_S(\Omega_S+\Omega_B)} \frac{\int{I_\nu d\nu}}{\theta_a\theta_b}\frac{Q(T_{ex})e^{E_u/T_{ex}}}{\nu^3 S_{ij}\mu^2} \mbox{ cm}^{-2}
\end{equation}
where $N_{tot}$ is the total column density [cm$^{-2}$] and $T_{ex}$ is the excitation temperature [K]. Line width and velocity were extracted from the Gaussian fits used in the $\int I_\nu d\nu$ term, which is the integrated intensity of the line [Jy beam$^{-1}$ {\kms}]. The optical depth correction factor $C_{\tau} = \tau/(1-e^{-\tau})$ was assumed to be unity, since the optical depth $\tau$ is known to be small for the isotopologue transitions selected (see Appendix~\ref{sec:LTE}). 
The $\Omega_S$ and $\Omega_B$ terms are the solid angles of the source and beam respectively; we assume that the source completely fills the beam at all positions and that $\Omega\approx\pi\theta^2$ since $\theta$ is small. 
The remaining terms are known: $\theta_a$ and $\theta_b$ are the FWHM beam sizes [\arcsec], $Q$ is the partition function, $E_u$ is the upper-state energy level [K], $\nu$ is the rest frequency of the transition [GHz], and $S_{ij}\mu^2$ is the product of the transition line strength and the square of the electric dipole moment [Debye$^2$].  

\begin{deluxetable*}{lccchcc}
\tablecaption{Transitions used for line fits.}\label{tab:lines}
\tablehead{\colhead{\multirow{2}{*}{Transition}} & \colhead{$\nu$} & \colhead{$E_u$} & \colhead{$S_{ij}\mu^2$} & & \colhead{$A_{ul}$} & \colhead{\multirow{2}{*}{$g_u$}}
\\ 
 & \colhead{(GHz)} & \colhead{(K)} & \colhead{(Debye$^2$)} & & \colhead{($\times 10^{-5}$ s$^{-1}$)} &  
 }
\startdata
\multicolumn{6}{l}{\textbf{\ce{^13CH3OH}}\tablenotemark{a}}\\
$8_{(0,8)}-8_{(-1,8)}$ & 155.6958 & 94.59 & 6.86 & -4.75147 & 1.77 & 17\\
$6_{(0,6)}-6_{(-1,6)}$ & 156.1866 & 60.66 & 5.70 & -4.71127 & 1.94 & 13\\
$5_{(0,5)}-5_{(-1,5)}$ & 156.2994 & 47.08 & 4.98 & -4.69628 & 2.01 & 11\\
\multicolumn{6}{l}{\textbf{\ce{CH3CN} $\nu_8=1$}\tablenotemark{b}}\\
$J=8_{-5}-7_{-5}$ & 147.5123 & 802.03 & 278.88 && 15.32 & 68\\
$J=8_{-4}-7_{-4}$ & 147.5439 & 724.50 & 171.63 && 18.87 & 34\\
$J=8_{-3}-7_{-3}$ & 147.5698 & 661.23 & 196.67 && 21.64 & 34\\
$J=8_{5}-7_{5}$ & 147.5756 & 668.87 & 139.43 && 15.34 & 34\\
$J=8_{-2}-7_{-2}$ & 147.5899 & 612.23 & 429.03 && 23.61 & 68\\
$J=8_{4}-7_{4}$ & 147.5954 & 618.01 & 343.26 && 18.89 & 68\\
$J=8_{-1}-7_{-1}$ & 147.6040 & 577.50 & 225.29 && 24.80 & 34\\
$J=8_{2}-7_{2}$ & 147.6199 & 559.01 & 214.52 && 23.62 & 34\\
\enddata
\tablenotetext{a}{Spectroscopic data for \ce{^13CH3OH} comes from CDMS.}
\tablenotetext{b}{Spectroscopic data for \ce{CH3CN} $\nu_8=1$ comes from the JPL spectroscopic database.}
\end{deluxetable*}

\begin{figure*}
\plotone{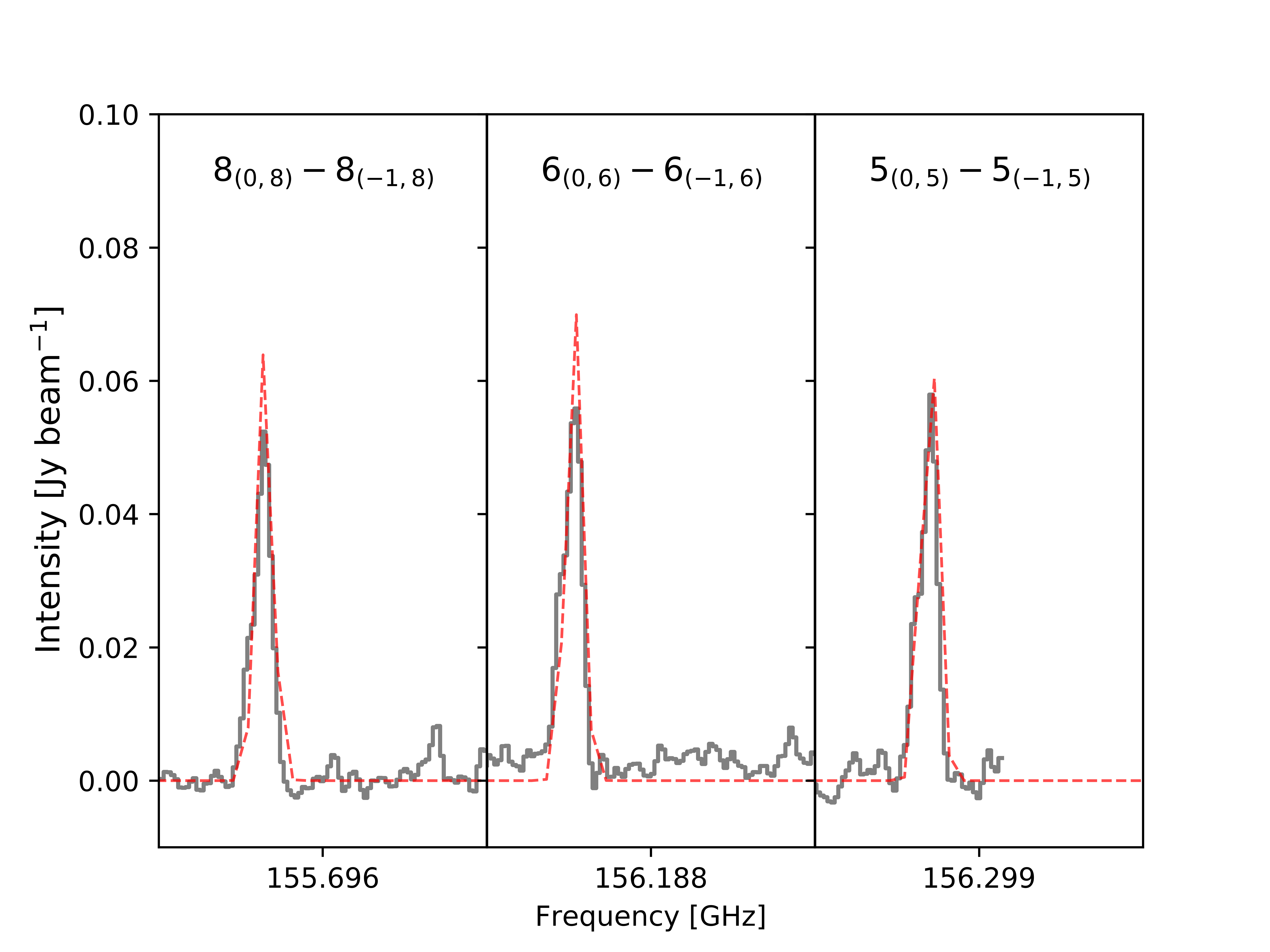}
\caption{{Example spectrum of \ce{^13CH3OH} $\nu=0$ toward the center of the Compact Ridge showing the Gaussian fits to the unblended transitions by the red dashed line over the data in black.}}
\label{fig:spectrum}
\end{figure*}

The transitions of \ce{^13CH3OH} and \ce{CH3CN} used for the fits are summarized in Table~\ref{tab:lines}. One other \ce{^13CH3OH} transition is present in our observations at 155.262 GHz but was not included in the fits due to line blending. The \ce{^13CH3OH} lines are ground vibrational state \textit{c}-type transitions with upper state energies $E_u$ between 47 and 95 K and $J=5,6,8$. 
The $J=7$ line at $\nu_{rest}=155.994$ GHz is expected to be unblended but was excluded in our observations to optimize the spectral setup. Three other \ce{^13CH3OH} transitions fall within our spectral setup at $\nu_{rest} = 141.381$ GHz, $155.262$ GHz, and  $156.149$ GHz. {The 141.381 GHz and 155.262 GHz lines appear to be blended: with \ce{CH3OH} $v_t=1$ at 141 GHz and an unknown contaminant at 155 GHz.} The line at 156.149 GHz is high $J$ ($J=26-25$) with $E_u = 988$ K and thus a strong detection of this transition would not be expected especially toward the Compact Ridge. An example spectrum of the \ce{^13CH3OH} lines fitted in our analyses is shown in Figure~\ref{fig:spectrum}.

\begin{figure*}
\plotone{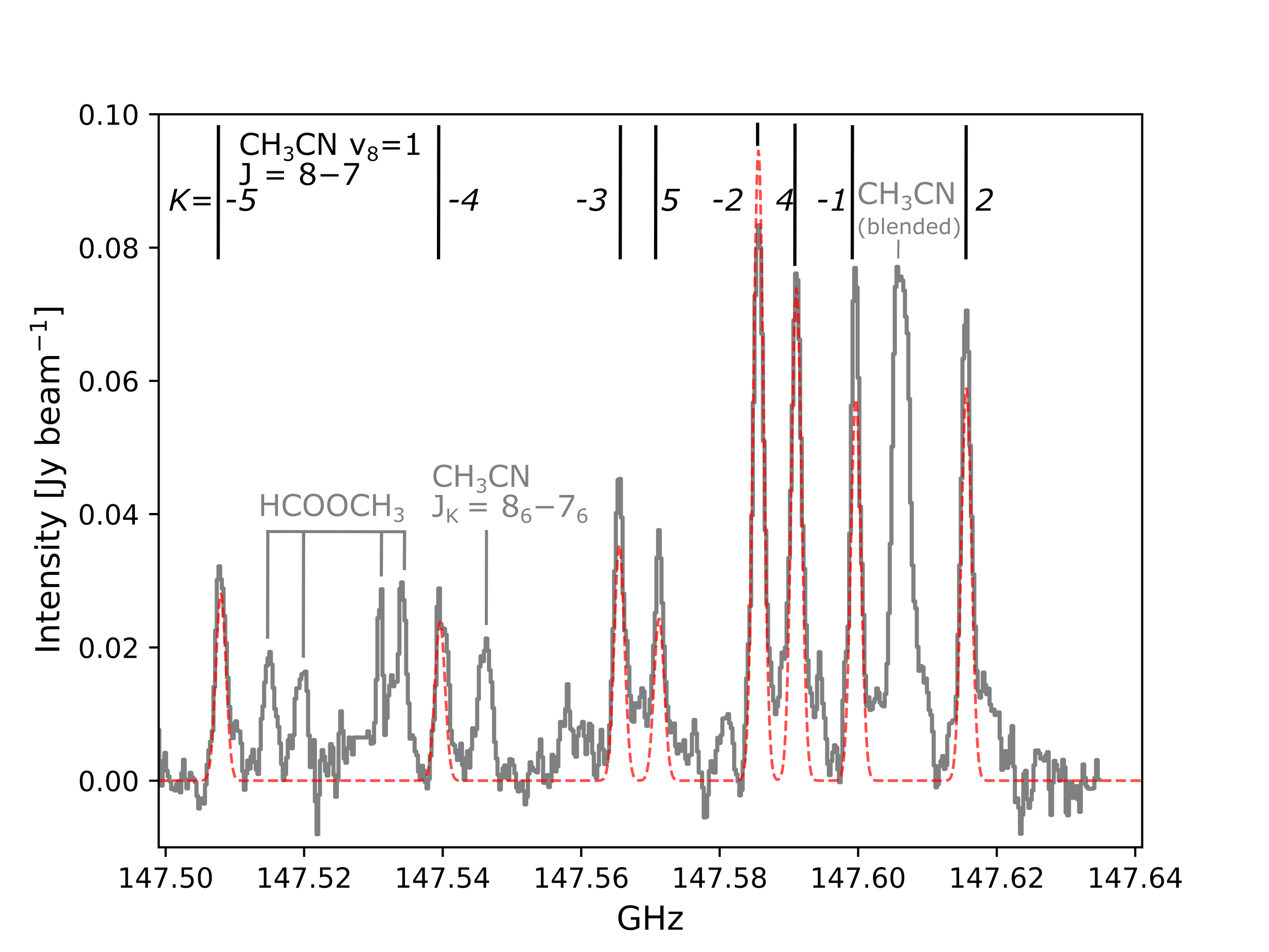}
\caption{{Example spectrum of \ce{CH3CN} $\nu_8=1$ toward the center of the Hot Core showing the Gaussian fits to the included transitions by the red dashed line over the data in black.}}
\label{fig:spectrumCH3CN}
\end{figure*}

The \ce{CH3CN} lines analyzed are all strong, vibrationally-excited ($\nu_8=1$) $J=8-7$ transitions. Their upper state energies between 559 K and 803 K make these lines appropriate tracers of high-energy (or radiatively pumped) emission, in and near the Hot Core. There are other, weaker \ce{CH3CN} $\nu_8=1$ lines in Figure~\ref{fig:spectrumCH3CN}; however, these lines were excluded from the fits because they {are blended, which prevented good line fits}. Nevertheless, the fitted parameters characterizing the \ce{CH3CN} emission come with {reasonable propagated uncertainties of $\le$20\% for much of the Hot Core (and up to $\sim$30\% near the region's edges where emission drops off)}.

Because the synthesized beams are small compared to the extent of molecular emission toward Orion KL, we assume the sources completely fill the beam at all positions. The $E_u$, $\nu$, $S_{ij}\mu^2$, and $Q(T_{ex})$ values for \ce{^13CH3OH} and \ce{CH3CN} $\nu_8=1$ were obtained from the Cologne Database for Molecular Spectroscopy \citep[CDMS,][]{Mueller2001} and the JPL molecular spectroscopy database \citep[][]{Pickett1998}, respectively, via the Splatalogue\footnote{\href{http://splatalogue.online}{http://splatalogue.online}} database for astronomical spectroscopy. Values and uncertainties (standard errors calculated using \texttt{LMFIT}) for $N_{tot}$, $T_{ex}$, and local standard of rest velocity $V_{LSR}$ were extracted from the line fits using Equation~\ref{eq:Ntot} and are presented throughout the rest of the paper.

\begin{figure}[t!]
\includegraphics[width=\textwidth]{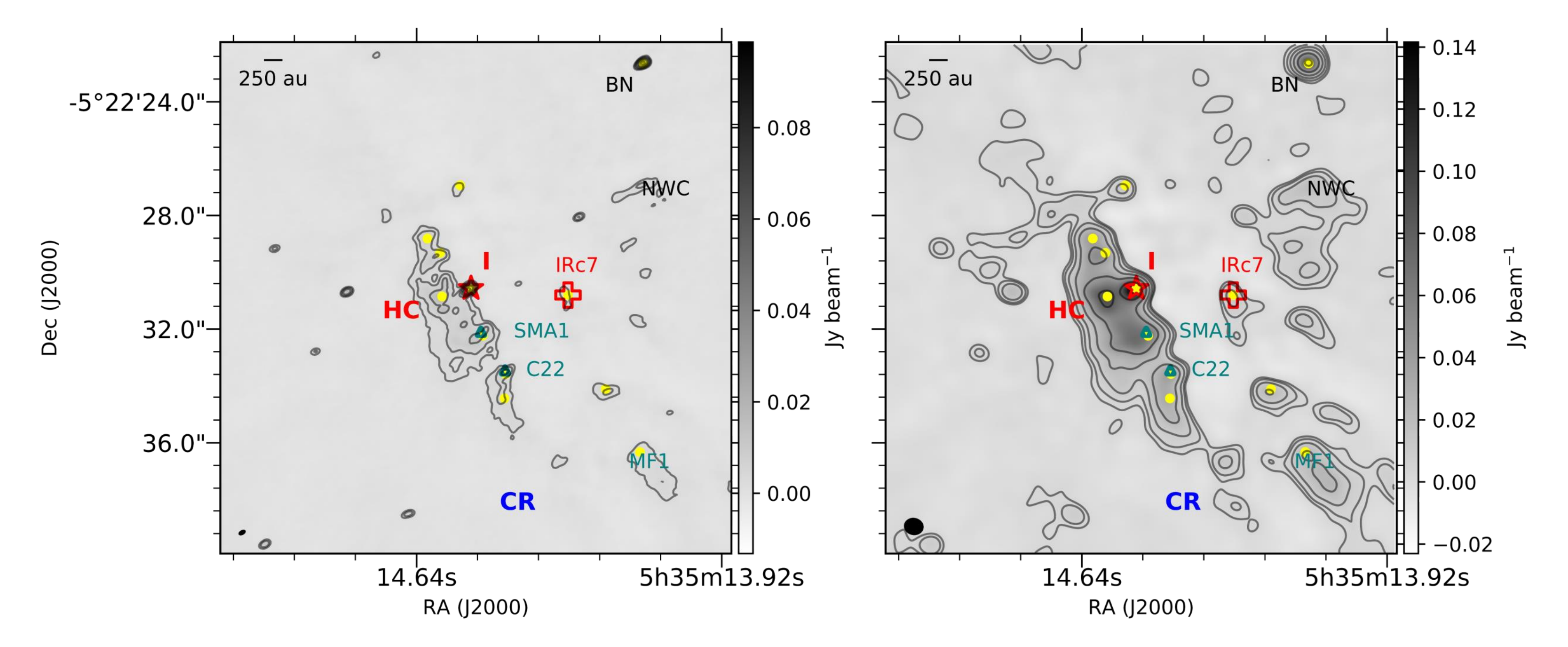}
\caption{ALMA Cycle 5 continuum in Band 4 ($\sim$150 GHz) toward Orion KL at full (left) and tapered (right) angular resolution. The synthesized beam is shown {by the black oval} in the bottom left. The warmer, denser Hot Core region is shown by \textcolor{red}{\textbf{HC}}, and the cooler, less dense Compact Ridge region is indicated by \textcolor{blue}{\textbf{CR}}. The locations of source I and IRc7 are shown by the red star and red plus, respectively. {The yellow circles correspond to millimeter/submillimeter continuum sources identified by \citet{Hirota2015}. The embedded millimeter source C22 \citep{Friedel2011} and the protostar SMA1 \citep{Beuther2006} are shown by the teal triangles.} Both the greyscale and contours plot the continuum with contours at $3\sigma_{\mbox{\scriptsize RMS}},6\sigma_{\mbox{\scriptsize RMS}},12\sigma_{\mbox{\scriptsize RMS}},24\sigma_{\mbox{\scriptsize RMS}},\dots$ where $\sigma_{\mbox{\scriptsize RMS}}=1.3$ mJy beam$^{-1}$ is the noise level for the full resolution image and $\sigma_{\mbox{\scriptsize RMS}}=1.0$ mJy beam$^{-1}$ is the noise level for the tapered resolution image.
}
\label{fig:continuum}
\end{figure}

\begin{figure}
    \centering
    \epsscale{1.1}
    \plotone{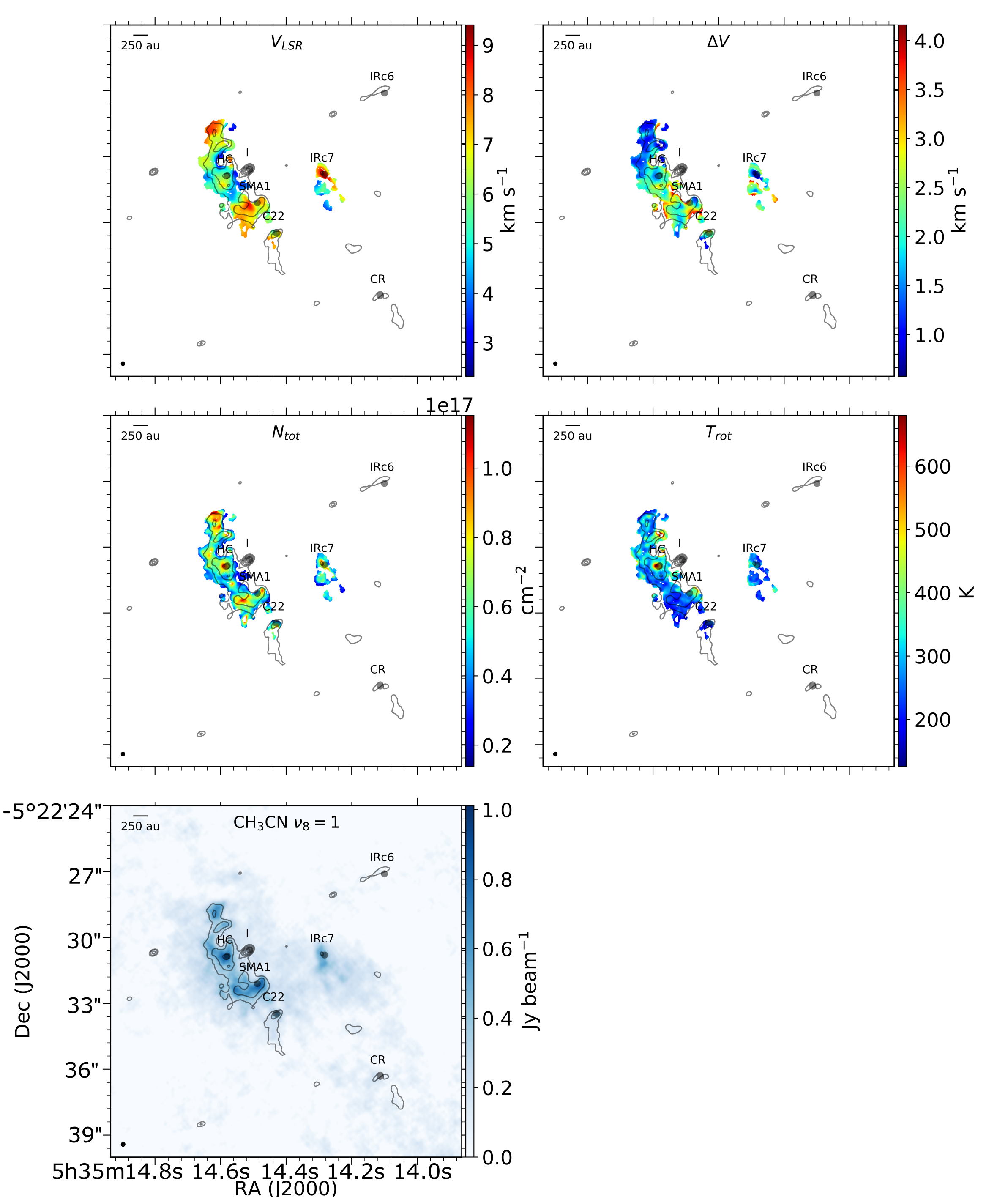}
    \caption{Parameter maps for the full angular resolution images of \ce{CH3CN} $\nu_8=1$. Contours show the 2 mm ($\sim$150 GHz) continuum from the left panel of Figure~\ref{fig:continuum}. {The color plots show, from left to right: (top row) velocity field, line width field, (middle row) total column density, excitation temperature, and (bottom row) integrated intensity (moment-0). All panels have the same field of view.}}
    \label{fig:resultshiresCH3CN}
\end{figure}

\begin{figure}
    \centering
    \epsscale{1.1}
    \plotone{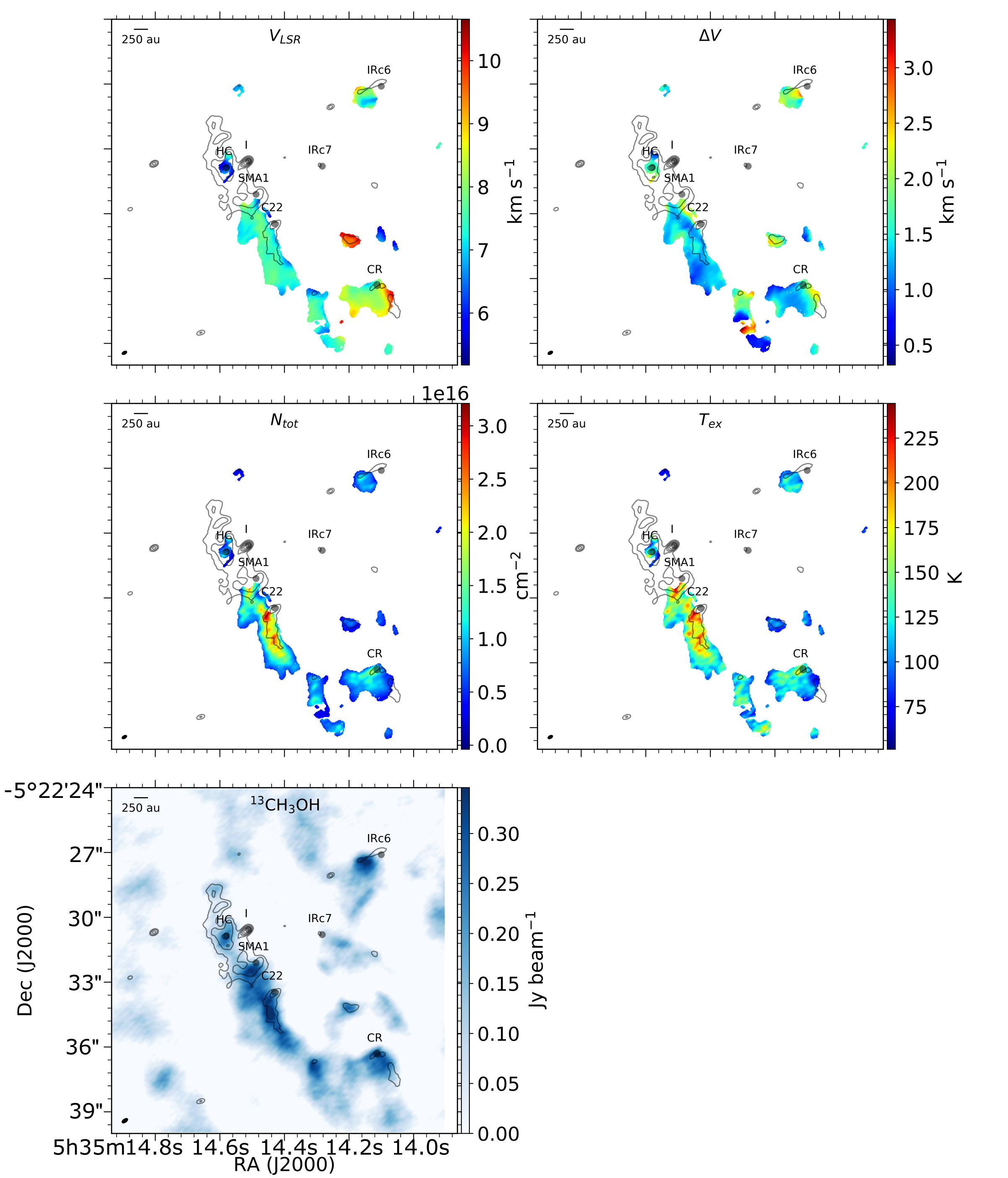}
    \caption{Parameter maps for the full angular resolution images of \ce{^13CH3OH} $\nu=0$. Contours show the 2 mm ($\sim$150 GHz) continuum from the left panel of Figure~\ref{fig:continuum}. {The color plots show, from left to right: (top row) velocity field, line width field, (middle row) total column density, excitation temperature, and (bottom row) integrated intensity (moment-0). All panels have the same field of view.}}
    \label{fig:resultshiresCH3OH}
\end{figure}

\begin{figure}
    \centering
    \epsscale{1.1}
    \plotone{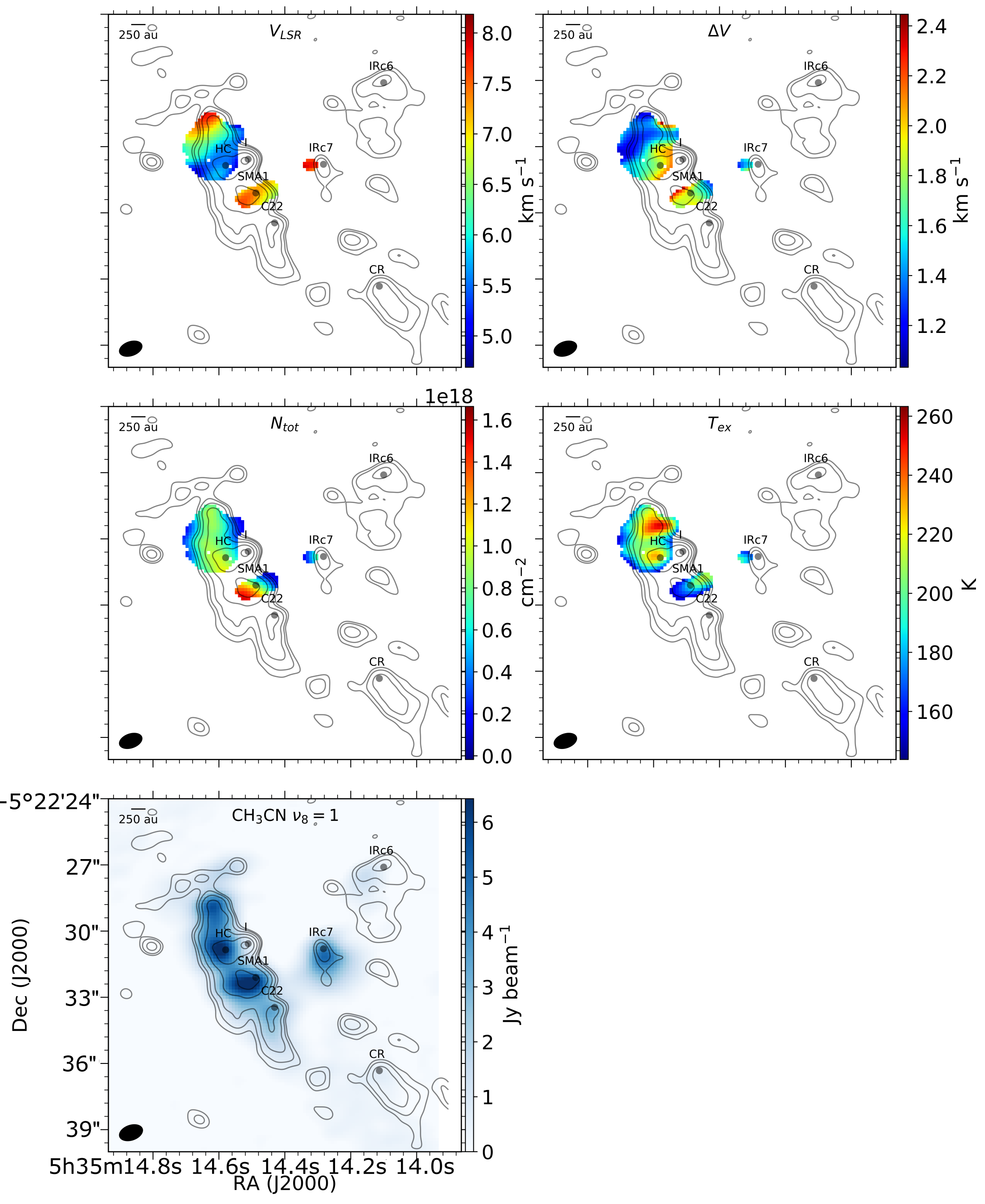}
    \caption{Parameter maps for the tapered angular resolution images of \ce{CH3CN} $\nu_8=1$. Contours show the 2 mm ($\sim$150 GHz) continuum from the right panel of Figure~\ref{fig:continuum}. {The color plots show, from left to right: (top row) velocity field, line width field, (middle row) total column density, excitation temperature, and (bottom row) integrated intensity (moment-0). All panels have the same field of view.}}
    \label{fig:resultslowresCH3CN}
\end{figure}

\begin{figure}
    \centering
    \epsscale{1.1}
    \plotone{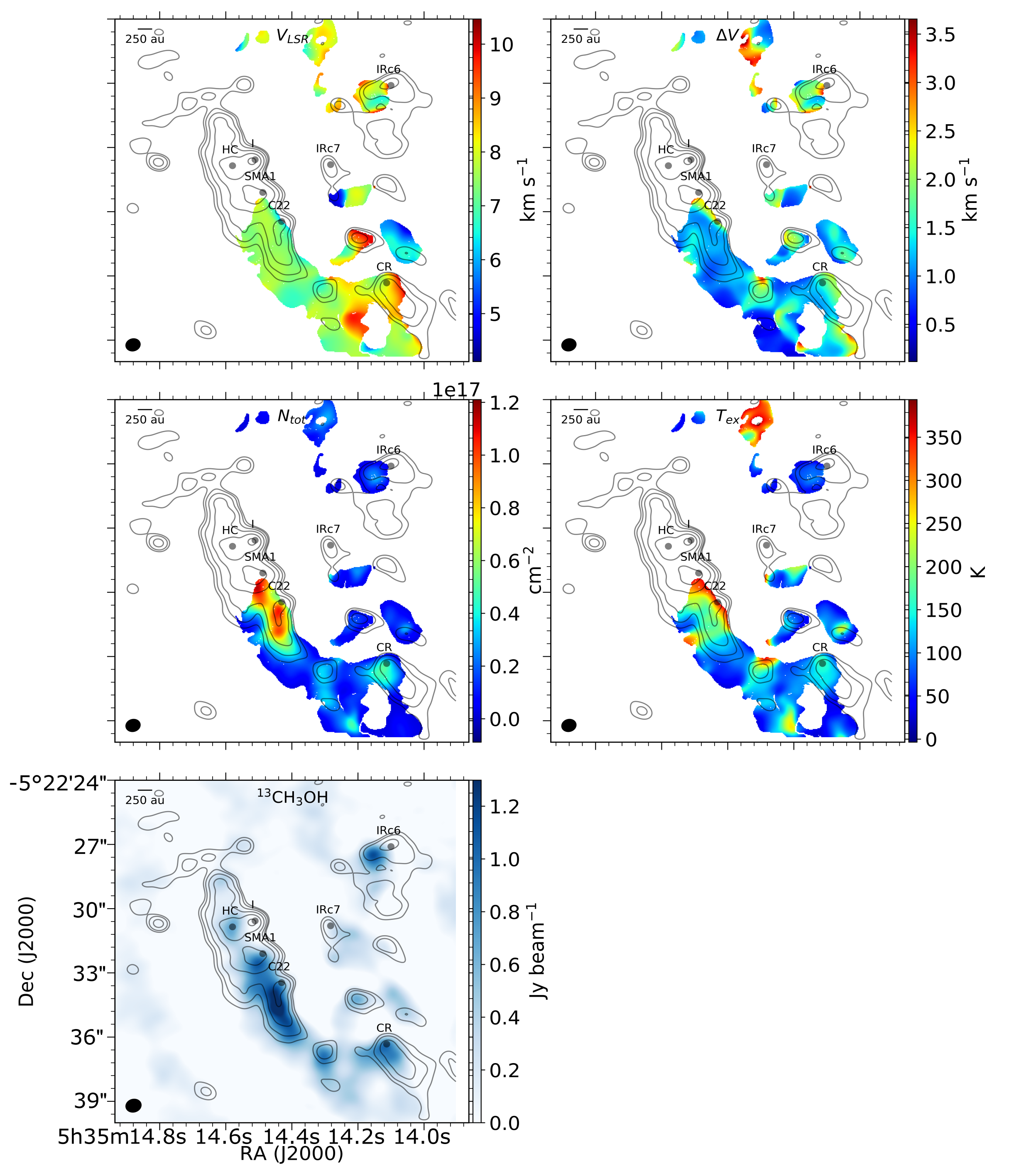}
    \caption{Parameter maps for the tapered angular resolution images of \ce{^13CH3OH} $\nu=0$ (bottom cluster). Contours show the 2 mm ($\sim$150 GHz) continuum from the right panel of Figure~\ref{fig:continuum}. {The color plots show, from left to right: (top row) velocity field, line width field, (middle row) total column density, excitation temperature, and (bottom row) integrated intensity (moment-0). All panels have the same field of view.}}
    \label{fig:resultslowresCH3OH}
\end{figure}

The figures presented herein show color maps of the derived parameters { from \ce{CH3CN} $\nu_8=1$ and \ce{^13CH3OH} $\nu=0$ line emission} with contours of the continuum at both the full and tapered angular resolution (Figures~\ref{fig:resultshiresCH3CN}-\ref{fig:resultslowresCH3OH}). In general, the estimated uncertainty is $<$30\% for the column density and excitation temperature and $<$5\% for the velocity and width fields. Maps of the propagated uncertainty are given in Appendix~\ref{sec:uncertainty}. The continuum contours are the same as those plotted over the Cycle 5 Band 4 ($\sim$150 GHz) continuum images in Figure~\ref{fig:continuum}. The continuum images obtained from the Cycle 4 data exhibit similar profiles and are not shown here. 

\section{Chemical Distributions}\label{sec:chemdist}

The { bottom-most panels of Figures~\ref{fig:resultshiresCH3CN}-\ref{fig:resultslowresCH3OH}} show the {integrated intensities} of \ce{^13CH3OH} and \ce{CH3CN} $\nu_8=1$ throughout Orion KL.  {While \ce{^13CH3OH} emission is present throughout the nebula, most of the emission arises from the region extending southwest from the Hot Core and toward the Compact Ridge, which fits the profile of this region being the site of oxygen-bearing chemistry \citep[e.g.][]{Blake1987,Favre2017,Tercero2018}.} \ce{CH3CN} $\nu_8=1$ is found predominantly in the region surrounding the Hot Core and Source I, as well as IRc7.

In our analyses, \ce{^13CH3OH} $\nu=0$ was selected to trace colder, less dense gas such as that in the Compact Ridge, a component that is also rich in oxygen-bearing chemistry \citep[][]{Blake1987,Friedel2011}. Furthermore, \ce{^13CH3OH} serves as an optically thin proxy for the primary carbon-12 isotopologue of methanol, which is often optically thick in star-forming regions and which is known to be optically thick toward the Hot Core and Compact Ridge in Orion KL specifically, from Herschel Space Telescope observations \citep[][]{Crockett2014}. Compared to lower angular resolution images \citep[e.g., CARMA images by ][]{Friedel2011}, the methanol emission in our observations is much more compact. We attribute this to the extended emission of the nebula being resolved out in our images. As stated in Section~\ref{sec:obscyc5}, we recover $\sim$9\% and $\sim$14\% of the \ce{^13CH3OH} flux in the 0\farcs3 and 0\farcs7 images compared to observations at 2\farcs0. Even between Figures~\ref{fig:resultshiresCH3OH} and ~\ref{fig:resultslowresCH3OH}, the (fittable) \ce{^13CH3OH} emission is generally more compact in the full angular resolution images (Figure~\ref{fig:resultshiresCH3OH}) than in the tapered angular resolution images (Figure~\ref{fig:resultslowresCH3OH}). 

The derived \ce{^13CH3OH} column densities are shown in the {middle} left panel Figures~\ref{fig:resultshiresCH3OH} and ~\ref{fig:resultslowresCH3OH}. We derive \ce{^13CH3OH} column densities on the order of $10^{16}$ cm$^{-2}$ throughout the Compact Ridge and the region southwest of the Hot Core (Hot Core-SW) at 0\farcs3, which is in agreement with the \ce{^13CH3OH} column densities reported by \citet{Crockett2014}. In the Compact Ridge and the edges of the Hot Core-SW, the column density is ${\sim}1.0 \times 10^{16}$ cm$^{-2}$; toward the center of the Hot Core-SW, there are pockets of higher column density, as much as ${\sim}3.0\times10^{16}$ cm$^{-2}$. At 0\farcs7 angular resolution, the derived column density is as much as four times higher, particularly in elongated structures extending south from SMA1 and C22. This is perhaps due to emission, which is resolved out in the 0\farcs3 images, from material that is being blown southward by the explosive event that took place north of Source I 500 years ago \citep[e.g., ][]{Bally2015}.

\ce{CH3CN} $\nu_8=1$ was selected for similar reasons, but to trace the hotter and denser gas toward the Hot Core region, which is also rich in nitrogen-bearing molecules \citep[][]{Blake1987,Friedel2011}. The vibrationally excited lines were specifically targeted to probe high energy emission excited by far-IR pumping associated with particularly dense gas from massive embedded protostars \citep[e.g.][]{Genzel1989} while also overcoming the fact that \ce{CH3CN} $\nu=0$ has been observed to be optically thick toward the Hot Core, Compact Ridge, and Plateau regions in Orion KL \citep[][]{Crockett2014}. 

The column density of \ce{CH3CN} $\nu_8=1$ is on the order of $10^{16}{-}10^{18}$ cm$^{-2}$ in the Hot Core and Source I region, with the column density derived from the 0\farcs9 data being about an order of magnitude greater than that derived from the 0\farcs 2 data. \citeauthor{Crockett2014} report \ce{CH3CN} $\nu_8=1$ column densities of $3.7\times 10^{15}$ cm$^{-2}$, which is almost two orders of magnitude less than what we generally find in our fits, toward the Hot Core in a $\ge$9\arcsec~beam. The discrepancy may be the result of probing vastly different spatial scales ($\ge$9\arcsec ~versus sub-arcsecond angular scales) or because \citeauthor{Crockett2014} use a different analytical method, including a wideband survey whereas our analyses includes $J=8$ transitions of \ce{CH3CN} $\nu_8=1$ only. We suspect that the former is the likely source of the difference between the two sets of measurements because \ce{CH3CN} $\nu_8=1$ traces dense gas, in particular gas affected by far-IR pumping associated with massive embedded protostars. In other words, we can expect the \ce{CH3CN} $\nu_8=1$ to be more compact such that its emission was perhaps diluted in a much larger beam. 

Alternatively, the $\nu_8=1$ mode is not necessarily in thermal equilibrium with the \ce{CH3CN} ground state at the same effective rotational temperature (i.e., $T_{vib} \neq T_{rot}$). However, \citeauthor{Crockett2014} report that the $\nu=0$ and $\nu_8=1$ models are consistent toward the Hot Core, or that $T_{rot} = T_{vib}$, and we find that the rotational LTE models employed in this work fit the data well (see Appendix~\ref{sec:LTE}). Thus we proceed under the assumption of thermal equilibrium, with the caveat that the derived excitation temperatures for the $\nu_8=1$ lines do not necessarily reflect the actual kinetic temperature. However, previous measurements for ground-state \ce{CH3CN} at $0\farcs33\times0\farcs35$ \citep[][]{Carroll2017thesis} yield similar temperature values as what we see in our 0\farcs9 data. These values are also similar to the temperature measurements in our 0\farcs2 data cube except at the very center of the Hot Core over an area presumably resolved out in the larger beam. 


\section{Physical Properties}\label{sec:structure}

{The (sub)mm continuum of Orion KL is mostly dominated by thermal dust emission \citep[e.g., ][]{Masson1985,Wright1985,Wright2017}; however, some regions---namely Source I---may have significant contributions from free-free emission \citep[][]{Beuther2006}}.  It is well-established that the Hot Core region is denser than the Compact Ridge, and this is evident by the greater thermal dust emission toward the Hot Core in our continuum images as well, as seen in Figure~\ref{fig:continuum}. The velocity maps derived here, consistent with previous results, show that the Hot Core and Source I regions have a lower LSR velocity than the Compact Ridge, one that is blue shifted with respect to the large scale molecular cloud complex, perhaps showing a profile resulting from the recent Orion KL explosive event. 

 The top {left} panel of each set of {Figures~\ref{fig:resultshiresCH3CN}-\ref{fig:resultslowresCH3OH}} show the velocity fields derived from \ce{^13CH3OH} $\nu=0$ and \ce{CH3CN} $\nu_8=1$ line emission. The velocity fields for each tracer were derived from the frequency offset in the pixel-by-pixel fits of the lines. Deriving the velocity offsets from a fixed frequency offset results in only a modest level of uncertainty because of Orion KL's low $V_{LSR}$, the high frequency of these observations, and the close proximity of the included lines (Table~\ref{tab:lines}): within 0.6 GHz for \ce{^13CH3OH} and 0.1 GHz for \ce{CH3CN}. This adds negligible uncertainty ($\ll1\%$) to the derived frequency offsets. This is confirmed by consistently low velocity uncertainties: the uncertainty on these measurements is ${\le}3.5\%$, except for a few regions in the 0\farcs2 \ce{CH3CN} map, in which some regions near the edges of the emission region have uncertainties as high as $\sim$8\%. Generally, the uncertainty is ${<}0.5$\% in all maps, and places with higher uncertainties often occur at the edges of emission regions and thus may be the result of the flux dropping off, giving rise to weak or noisy lines. 

The velocity gradients mapped by \ce{^13CH3OH} in this work agree well with past methanol observations, as summarized in Table~\ref{tab:velocity}. In brief, our derived velocity field in the Compact Ridge spans $\sim$7-9 \kms, and previous measurements have yielded an LSR velocity of $7.6{-}8.6$ \kms. 

\begin{deluxetable}{rc|c}
\tablecaption{Velocity profiles traced by molecular emission in Orion KL.}
\label{tab:velocity}
\tablehead{\colhead{\multirow{2}{*}{Molecule}} & \colhead{$V_{LSR}$} & \colhead{\multirow{2}{*}{Reference}} \\ 
 & \colhead{(\kms)} & 
}
\startdata
\multicolumn{3}{c}{Compact Ridge}\\\hline
\ce{^13CH3OH} & ${\sim}7{-}9$ & This work\\
\ce{^13CH3OH} & 7.8 & \citet{Crockett2014}\\
\ce{CH3OH} & 8.6 & \citet{Wang2011}\\
\ce{CH3OH} & 7.6 & \citet{Wilson1989}\\\hline
\multicolumn{3}{c}{Hot Core}\\\hline
\ce{^13CH3OH} & ${\sim}6$ & This work\\
\ce{^13CH3OH} & 7.5 & \citet{Crockett2014}\\
\ce{CH3OH} & $6{-}9$& \citet{Wang2011}\\
\ce{CH3OH} & ${\sim}7$& \citet{Wilson1989}\\
\ce{CH3CN} $\nu_8=1$ & ${\sim}6{-}8$ & This work\\
\ce{CH3CN} $\nu_8=1$ & $6{-}9$ & \citet{Wang2011}
\enddata
\end{deluxetable}

Our derived velocity toward the Hot Core region is slightly lower at $\sim$6 \kms ~in the \ce{^13CH3OH} images. Likewise, the LSR velocity derived from \ce{CH3CN} $\nu_8=1$ spans $\sim$6-8 \kms, which agrees with \citet{Wang2011}, who report a $V_{LSR}$ of 6-9 \kms ~toward the Hot Core.

The velocity field provides insight into the physical structure of Orion KL at spatial scales of ${\le}350$ au. Specifically, we find possible evidence of a molecular outflow in the Hot Core and a velocity gradient in the Hot Core-SW. 

\subsection{Hot Core}\label{sec:structureHC}

\begin{figure}
    \centering
    \epsscale{1.0}
    \plotone{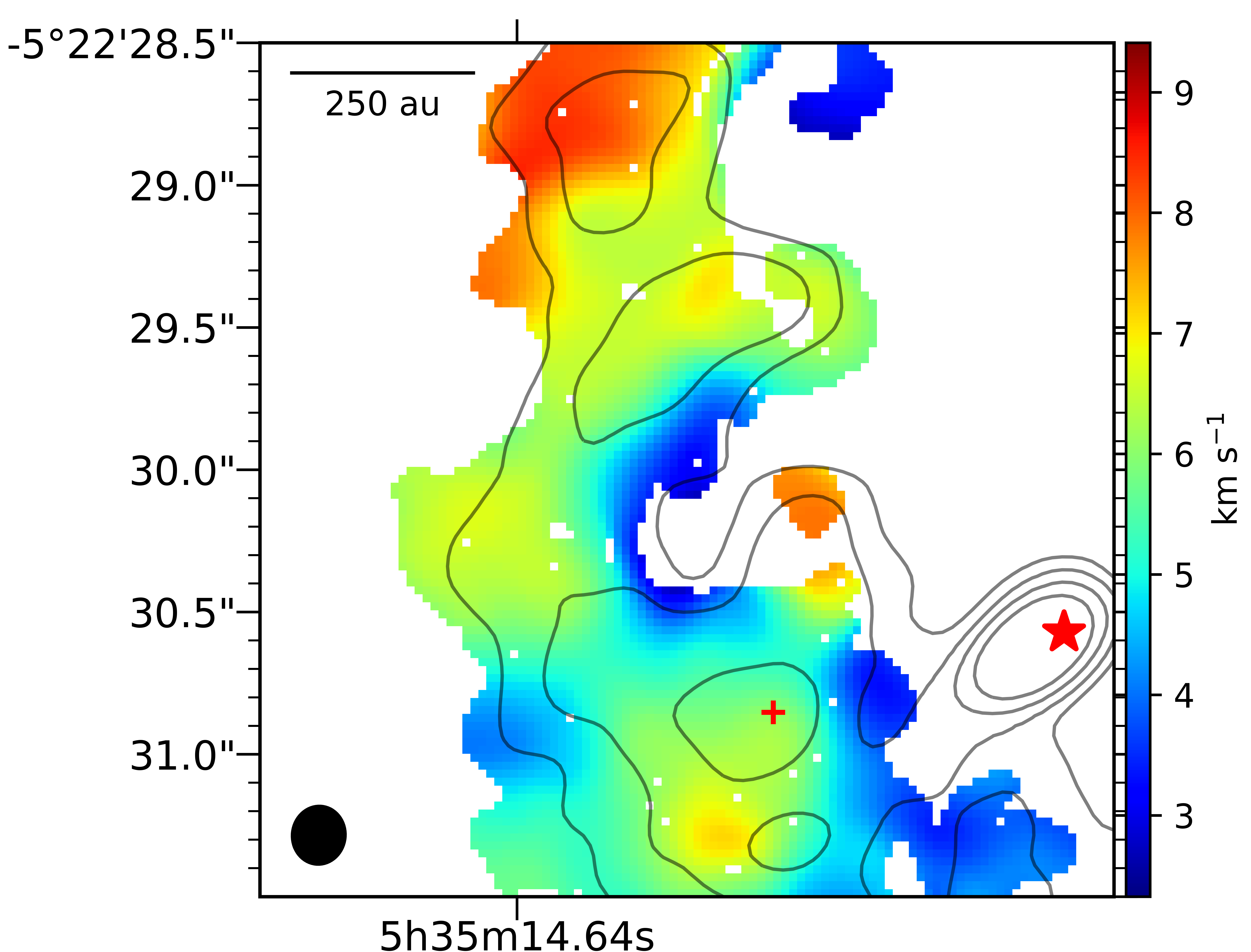}
    \caption{The same as the \edit1{\ce{CH3CN} $\nu_8=1$} {$V_{LSR}$ (top left) panel of Figure~\ref{fig:resultshiresCH3CN}} but with a close-up view of the Hot Core region. The red cross shows the mm/submm emission peak, and the red star denotes Source I. {Contours show the 2 mm ($\sim$150 GHz) continuum from the left panel of Figure~\ref{fig:continuum}.}}
    \label{fig:velocityzoom}
\end{figure}

In our \ce{CH3CN} 0\farcs2 velocity map, the Hot Core mm/submm emission peak \citep[alternatively designated SMM3 by][]{Zapata2011} is characterized by a radial velocity gradient that runs from $\sim$5 \kms ~(the LSR velocity of Source I) around the edge of its signature heart shape to values closer to 7-8 \kms to both the north and south of Source I. This velocity gradient is centered about 0.45\arcsec ~to the south of the Hot Core's 865 $\mu$m emission peak, as seen by the light orange region near the center of Figure~\ref{fig:velocityzoom}. This contrasts with the location of the peak excitation temperature and column density of \ce{CH3CN}, which are centered approximately at the location of the 865 $\mu$m emission peak. Opposite the Hot Core mm/submm peak, to the north, the velocity approaches $\sim$8 \kms.

Restated in terms of 3-D structure, the velocity field in the north is close to the systemic velocity of Orion KL \citep[9 \kms, e.g.,][]{Hall1978} but becomes more blue-shifted for the emission to the south associated with the Hot Core and Source I. With increasing blue shift, the velocity width also increases, as is seen in the top right panel of Figure~\ref{fig:resultshiresCH3CN}. The combination of the relative blue shift of the gas and the broader line widths may indicate that this dense gas is being pushed away from the putative explosive event. Areas throughout the Hot Core and Source I region where the velocity is closer to the systemic velocity, then, may be acting as a buffer to slow the gas, whereas a lower LSR velocity shows where that gas is pushed outward with fewer obstacles.


Another explanation for the elongated velocity structure is that it may be the result of a low-velocity outflow emanating from a source east of Source I. \citet{Goddi2011} detect a similar elongated velocity structure toward IRc7 in Expanded Very Large Array (EVLA) observations of high-excitation \ce{NH3} lines in a 0\farcs8 beam. They conclude that the elongated velocity structure, coupled with line-width enlargements, is evidence of a low-velocity outflow. As seen in the bottom right panel of {Figure ~\ref{fig:resultshiresCH3CN}}, \ce{CH3CN} $\nu_8=1$ has narrower line widths ($\sim$ 1.5 \kms) at the location of the Hot Core mm/submm peak, but broader lines ($\sim$3 \kms) in the region surrounding the peak. The fact that the mm/submm peak is co-spatial with a region of relatively narrow line widths and peaks in \ce{CH3CN} excitation temperature and column density but the elongated velocity structures are co-spatial with broader lines supports the conclusion that the elongated velocity structure to the north and south of this mm/submm peak is the result of a low-velocity outflow.  

\subsection{SMA1}\label{sec:structureSMA1}

SMA1 is a submillimeter source first identified by \citet{Beuther2004} at 346 GHz (865 $\mu$m). It has since been observed to have a 3 mm emission peak and a \ce{HCOOCH3} emission peak \citep[][]{Friedel2011,Favre2011}. \citet{Beuther2008} suggest that SMA1 hosts the driving source of the high-velocity outflow that runs northwest-southeast in Orion KL, but the nature of this object otherwise remains not well-understood. 

The structure surrounding SMA1 is particularly dynamic in our images. The velocity fields derived from the \ce{CH3CN} $\nu_8=1$ emission line at both 0\farcs 2 and 0\farcs9 angular resolution show a gradient that seems to emanate from just southeast of the SMA1 submm emission center. As seen in edit1{Figure~\ref{fig:resultshiresCH3CN}}, the velocity peaks at around 8 \kms. The velocity gradient decreases gradually moving to the southeast (seen in the 0\farcs2 map only) and northwest but changes abruptly in the orthogonal direction. This agrees with previous conclusions that SMA1 is the location of a driver of the northwest-southeast outflow. 

\subsection{Hot Core-SW}\label{sec:structureHCSW}

The Hot Core-SW, the region to the southwest of the Hot Core and east of the Compact Ridge, is distinct from, but shares characteristics with, both the Hot Core and the Compact Ridge. The velocity fields in this region, traced by \ce{^13CH3OH}, show a gradient from lower LSR velocity toward SMA1 ($\sim$6.5 \kms) to higher velocities ($\sim$7.8 \kms) toward the south near the Compact Ridge. This gradient can be explained by the interaction between gas emanating from the center of the explosion northwest of Source I with the clumps distributed throughout the nebula. As the gas travels away from the explosion center toward the east and south, it first interacts with the denser Hot Core and Source I regions, which act as a buffer to slow the gas. Toward the Compact Ridge in the south, there is less dust to act as a buffer, allowing the gas to pass through more easily and thus become more blue-shifted.

In this sense, the Hot Core-SW region is like the Hot Core and Source I in that it is denser than the Compact Ridge. Moreover, the Hot Core-SW contains higher column densities of \ce{^13CH3OH} than the Compact Ridge, a characteristic that is evident in both the 0\farcs3 and 0\farcs7 maps. This region also has higher excitation temperatures than the Compact Ridge. 

It is curious, then, that the Hot Core-SW region is not traced by nitrogen-bearing compounds, which is evident here from the lack of \ce{CH3CN} $\nu_8=1$ emission and from previous observations \citep[e.g.,][]{Friedel2008}. Despite having velocity, density, and temperature profiles more akin to the Hot Core region, the Hot Core-SW region appears to be more chemically similar to the Compact Ridge. 

Previous interferometric observations show strong emission in the Hot Core-SW from the oxygen-bearing species dimethyl ether \citep[\ce{(CH3)2O}, CARMA,][]{Friedel2008} and methyl formate \citep[\ce{HCOOCH3}, IRAM PdBI,][]{Favre2011}. \citet{Friedel2008} also found acetone (\ce{(CH3)2CO}) emission toward the Hot Core-SW, part of which overlaps with ethyl cyanide (\ce{C2H5CN}) emission near SMA1 and C22 in the north. In their observations, \ce{(CH3)2CO} was detected only toward the Hot Core-SW and IRc7, both of which have overlapping oxygen- and nitrogen-bearing chemistry. Subsequent observations of \ce{(CH3)2CO} show that this molecule is much more extended than reported by \citeauthor{Friedel2008} and that it is an oxygen-bearing species that has a distribution similar to that of large nitrogen-bearing molecules \citep[][]{Peng2013}. 

It would appear, then, that the Hot Core-SW is a hybrid of the Hot Core and Compact Ridge regions, and thus the Hot Core-SW presents itself as a possible location for bridging the oxygen-bearing chemistry of the less dense, colder Compact Ridge and the nitrogen-bearing chemistry of the denser, warmer Hot Core region.  {\citet{Sutton1995} suggest that the chemical differentiation between the Hot Core and Compact regions is not so stark as originally thought and that the chemical emission from both nitrogen-bearing and oxygen-bearing species is spatially extended.  Nevertheless, such chemical differentiation has also been observed in other sources, such as the ultracompact H\;\small{II} region W3(OH) \citep[][]{Wyrowski1997}, the high-mass protostar AGAL 328.25 \citep[][]{Csengeri2019}, and cluster-forming clumps in NGC 2264 \citep{Taniguchi2020}. In a sample of seven high-mass protostars, \citet{Bisschop2007} found that complex organics could be classified as either cold ($<$100 K) or hot ($>$100 K) based on their rotation temperatures, and that nitrogen-bearing species were exclusively ``hot'' molecules whereas different oxygen-bearing species were found in both categories. \citet{Fayolle2015} found chemical differentiation in a sample of three massive young stellar objects (YSOs); in their sample, \ce{CH3CN} and \ce{CH3OCH3} originated from the central cores, \ce{CH3CHO} and \ce{CH3CCH} emission came from the extended envelope, and \ce{CH3OH} (and sometimes \ce{HNCO}\edit1{)} straddled both regimes.  While the origin(s) of this chemical differentiation are not well understood, it may may be influenced by the precursor content (e.g., \ce{H2O}, \ce{NH3}, \ce{CH4}) in icy grain mantles \citep[e.g., ][]{Fayolle2015} and prestellar core evolutionary time-scales \citep[e.g., ][]{Laas2011,Sakai2013}. In Orion KL, this may translate to a scenario in which, on large spatial scales, the Hot Core-SW is analogous to the region between an embedded YSO core and its extended envelope, which would begin to explain the observed distribution of \ce{^13CH3OH} in this work and \ce{(CH3)2CO} by \citet{Friedel2008}. However, a better understanding of complex organic formation---and of the initial conditions in Orion KL before the putative explosive event---is likely needed to explain the observed chemical and physical patterns.}

\section{Thermal Structure}\label{sec:thermalstructure}

In considering the thermal structure of Orion KL, we assume local thermodynamic equilibrium (LTE, see Appendix~\ref{sec:LTE}) and that the kinetic temperature in the nebula is approximated by the excitation temperature of the molecular gas due to the nebula's high density. The derived temperature maps for \ce{^13CH3OH} $\nu=0$ and \ce{CH3CN} $\nu_8=1$ are shown in the {middle} right panels of each group of plots in Figures ~\ref{fig:resultshiresCH3CN} and ~\ref{fig:resultslowresCH3OH}, respectively. Because we have temperature maps on two different angular scales, we can get information about possible sources of heating throughout the nebula. {In particular, the higher angular resolution ($\sim$0\farcs2$-$0\farcs3) images probe spatial scales on the order of $\sim$78$-$116 au, which is within the protostellar radius of 150 au where the gas-phase chemistry is expected to be driven by thermal submlimation from the young stellar object \citep[e.g., ][]{Schoeier2002}. Conversely, the tapered angular resolution ($\sim$0\farcs7$-$0\farcs9) images correspond to spatial resolutions of $\sim$272$-$350 au, which is greater than the typical radius of a protostellar core. That is, our angular resolutions enable us to look at possible heating sources on scales that straddle the boundary of whether thermal sublimation is expected to be driven by an embedded protostar.}

We consider possible sources of heating in Orion KL by mapping the difference between the excitation temperatures derived in a larger versus smaller beam. That is, we subtract the excitation temperature {derived within} the smaller beam $T_{ex,S}$ from that derived {within} the larger beam $T_{ex,L}$ to give $\Delta T_{ex} = T_{ex,L}-T_{ex,S}$ {at each pixel}. It follows then that $\Delta T_{ex}>0$ suggests an external source of heating whereas $\Delta T_{ex}<0$ suggests an internal source of heating. {In other words, $\Delta T_{ex}<0$ indicates that the warmer temperature, potentially from a compact heating source, measured in a smaller beam is diluted by colder temperatures in the more extended larger beam. Similarly, $\Delta T_{ex}>0$ indicates that a colder region detected in a smaller beam is in the presence of nearby warmer gas.}

We note that whether the sign of $\Delta T_{ex}$ accurately reflects the source of heating depends on the assumed source radius and proximity to the source peak. As such, we ran simple Gaussian temperature distribution models for externally and internally heated material with approximate temperatures and radii of sources described in this section. In these toy models, briefly described in Appendix~\ref{sec:toymodels}, the sign of $\Delta T_{ex}$ consistently reflected whether a modeled compact source was internally or externally heated. {Furthermore, we note that, as mentioned in Section~\ref{sec:obscyc5}, we recover about the same amount of \ce{^13CH3OH} flux in both the full and tapered image cubes, indicating that $\Delta T_{ex}$ is not affected by flux recovery in the 0\farcs3 and 0\farcs7 synthesized beams.}

\begin{figure}[p]
    \centering
    \includegraphics[width=1.05\textwidth]{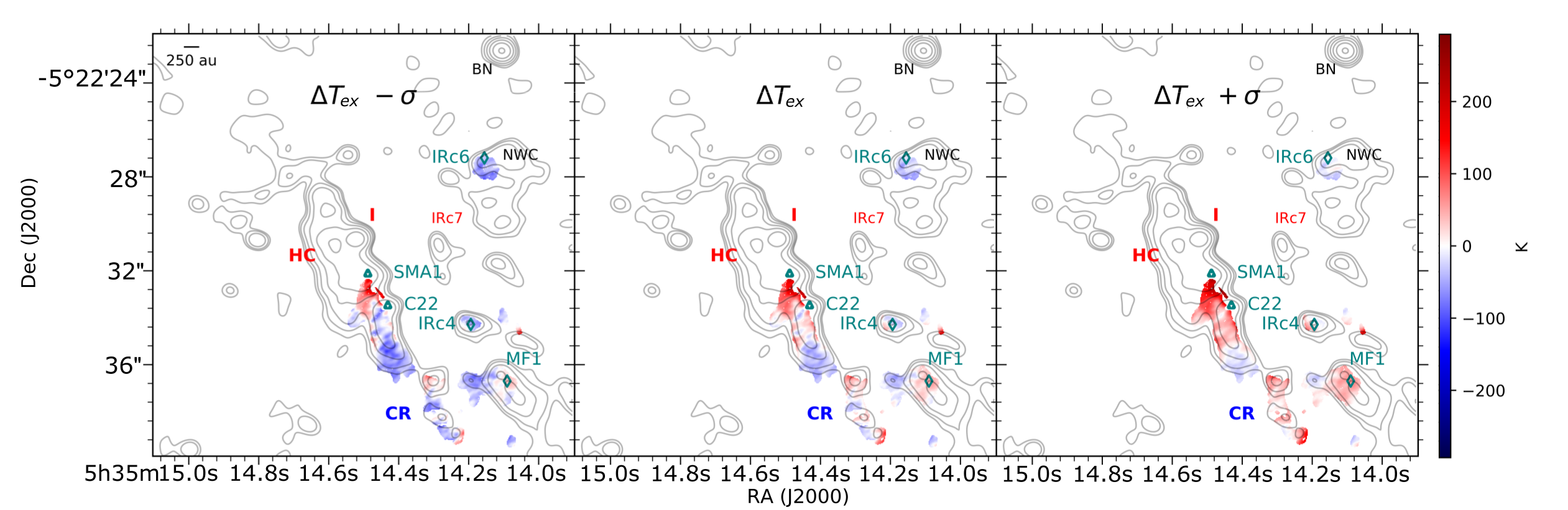}
    \caption{Maps showing $\Delta T_{ex}=T_{ex,0\farcs7}-T_{ex,0\farcs3}$, i.e., the difference between the derived excitation temperature from a 0\farcs7 and 0\farcs3 beam for \ce{^13CH3OH}. From left to right, the panels show $\Delta T_{ex}-\sigma$, $\Delta T_{ex}$, and $\Delta T_{ex}+\sigma$ where $\sigma$ is the propagated uncertainty. Red represents an external source of heating, and blue represents internal heating. {Contours show the 2 mm ($\sim$150 GHz) continuum.}}
    \label{fig:heatingsourceCH3OH}
\end{figure}
\begin{figure*}[p]
    \centering
    \includegraphics[width=1.05\textwidth]{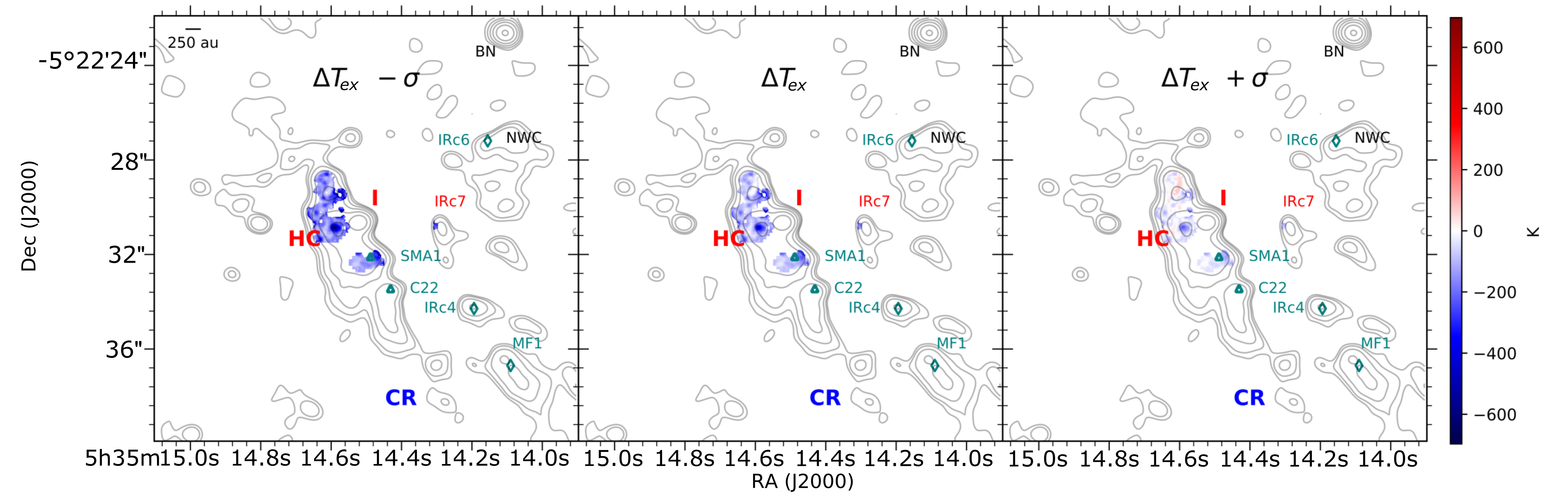}
    \caption{Maps showing $\Delta T_{ex}=T_{ex,0\farcs9}-T_{ex,0\farcs2}$, i.e., the difference between the derived excitation temperature from a 0\farcs9 and 0\farcs2 beam, for \ce{CH3CN} $\nu_8=1$. From left to right, the panels show $\Delta T_{ex}-\sigma$, $\Delta T_{ex}$, and $\Delta T_{ex}+\sigma$ where $\sigma$ is the propagated uncertainty. Red represents a possible external source of heating, and blue represents possible internal heating. {Contours show the 2 mm ($\sim$150 GHz) continuum.}}
    \label{fig:heatingsourceCH3CN}
\end{figure*}

The resulting maps for $\Delta T_{ex}$ are presented in Figures~\ref{fig:heatingsourceCH3OH} and \ref{fig:heatingsourceCH3CN} for \ce{^13CH3OH} and \ce{CH3CN} $\nu_8=1$, respectively. In these maps, regions where $\Delta T_{ex}>0$, indicative of an external source of heating, are shaded red. Regions where $\Delta T_{ex}<0$, indicative of a possible internal heating source, are shaded blue.

\subsection{Hot Core}\label{sec:heatingsourcesHC}

Our analyses suggest possible internal sources of heating in the Hot Core. As seen in Figure ~\ref{fig:heatingsourceCH3CN}, $\Delta T_{ex} < 0$ at the mm/submm emission peak of the Hot Core. We also see \ce{^13CH3OH} in the 0\farcs3 images only and not in the 0\farcs7 images, implying a highly compact methanol source, which would also be consistent with an internally heated core. This, along with a possibility of a low-velocity outflow, as discussed in Section~\ref{sec:structureHC}, matches the profile of an embedded young star.

Sources of heating in the Hot Core region {of Orion KL} remain the topic of ongoing debate. {In a more general sense, the term ``hot core'' describes a compact, molecule-rich region with temperatures between 100 and 300 K resulting from the warm-up of young massive stars \citep[e.g., ][]{Evans1999,Kurtz2000,Herbst2009}. Moreover, the nearly two dozen hot cores presented in the literature typically are characterized as being internally heated by embedded massive stars \citep[][]{HernandezHernandez2014,Fayolle2015}. However, there have been a couple of hot molecular cores---namely the Orion KL Hot Core and G34.26$+$0.15 \citep[][]{Mookerjea2007}---for which definitive evidence of an embedded heating source has eluded observations. Despite the expectation that a region of hot molecular gas would have an internal source of heating, the source of heating in Orion KL's Hot Core region is currently unknown. }

The so-called Hot Core was first identified by \citet{Ho1979}, who observed it as a compact source of hot \ce{NH3} emission. \citet{Blake1996} found no evidence for a (strong) source of internal heating there, and since then, there have been multiple studies that support an externally-heated ``hot core'' \citep[e.g.][]{Goddi2011,Zapata2011,Bally2017,OrozcoAguilera2017,Peng2017}. \citet{Zapata2011}, for example, favor an externally heated Hot Core model (SMM3 in their work) because they do not find hot molecular gas emission associated with self-luminous submillimeter, radio, or infrared sources. Furthermore, they find that hot \ce{HC3N} gas seems to form a shell around the northeastern edge of the Hot Core, which suggests the Hot Core is illuminated from the edges---that is, externally heated, specifically by shockwaves.

Alternatively, \citet{Kaufman1998} show that large, warm columns of molecular gas are more easily produced in internally heated cores. They conclude that, while Source I may externally heat some of the material in the area surrounding the Hot Core, there are likely young embedded star(s) heating the core. \citet{deVicente2002} further suggest that the Hot Core is internally heated by an embedded massive star obscured by an edge-on disk. Their speculation is based on vibrationally excited \ce{HC3N} observed in the Hot Core at energy levels that would require a heating source with too large a luminosity to come from nearby external sources. Furthermore, \citet{Crockett2014H2S} conclude that \ce{H2S} emission, which is pumped by far-IR radiation, in the Hot Core region signals the presence of a hidden source of luminosity. Another hybrid hypothesis is that the Hot Core might have been a typical hot core around Source I prior to Orion's explosive event, and that, as a result of this, even the dense gas of the Hot Core were spatially separated from the protostars, especially Source I \citep[][]{Nickerson2021}, giving us the enigmatic molecular core we observe today. 

We stress that our evidence in support of an internal heating source does not preclude external heating in the Hot Core. Other studies that have looked at whether the Hot Core is externally or internally heated have made use of a range of angular resolutions. For example, \citeauthor{Zapata2011} used Submillimeter Array (SMA) observations with an angular resolution of 3\farcs28$\times$3\farcs12, which corresponds to spatial scales of $>$1200 au in Orion KL. They suggest that external heating of the Hot Core comes from energetic shocks from the Orion KL explosion that are now passing through the Hot Core region. \citet{Goddi2011} drew similar conclusions but with EVLA observations at sub-arcsecond angular resolutions. 

Heating from a combination of shocks and embedded young stars may explain the line width profiles observed for \ce{CH3CN} $\nu_8=1$. At 0\farcs2 angular resolution, the \ce{CH3CN} $\nu_8=1$ line widths are narrower, with $\Delta V \sim1.5$ \kms, at the mm/submm Hot Core peak and are surrounded by \ce{CH3CN} $\nu_8=1$ emission marked by slightly broader lines of $\Delta V \sim 2.0$ \kms. While this is only a modest difference, the fact that the line width field is not uniform, specifically at the mm/submm peak, and that shocked material is marked by broader line widths are evidence that the conditions at this site are distinct from the surrounding region. 

\citet{Crockett2014} demonstrate that, relative to other components, the Hot Core is the most heterogeneous thermal structure in Orion KL given the spread of excitation temperatures derived for a spread of cyanides, sulphur-bearing molecules, and oxygen-bearing complex organics. Therefore, it is likely that the Hot Core region is heated by a multitude of sources, both internally and externally. Based on mass estimates of $1001\pm791 \;M_\odot$ \citep[][]{Pattle2017} and a 10\% star formation efficiency combined with a mean stellar mass of ${\sim}2 \;M_\odot$ \citep[][]{Hillenbrand1998}, the Orion KL/BN region could be expected to have enough mass to form $50\pm40$ stars. In other words, Orion is a dense protostellar cluster in which additional protostellar sources beyond Source I are expected. Further observations are needed to confirm whether there is, in fact, an embedded massive star at the Hot Core site as predicted by \edit1{\citet{deVicente2002}} and supported by our analyses.

\subsection{Compact Ridge}

Figure~\ref{fig:heatingsourceCH3OH} shows that the Compact Ridge, as traced by \ce{^13CH3OH} emission, is externally heated. This conclusion agrees with past studies of ground-state methanol in Orion KL \edit1{\citep{Wang2011}}. The derived excitation temperature for a \ce{HCOOCH3} emission peak in the Compact Ridge---denoted by the teal diamond labeled `MF1' in Figure~\ref{fig:heatingsourceCH3OH}---was also measured to be higher in a larger beam than in a smaller one \citep[][]{Favre2011}.

\subsection{Hot Core-SW}

As discussed in Section~\ref{sec:structureHCSW}, the region southwest of the Hot Core is interesting because it seems to exhibit, broadly speaking, the physical parameters of the Hot Core but the chemical composition of the Compact Ridge. The thermal structure of this region is also interesting because of its heterogeneity. 

The average \ce{^13CH3OH} excitation temperature in the Hot Core-SW (i.e., the mean derived $T_{ex}$ across all pixels in the range of $\alpha_{\mbox{\scriptsize J2000}} = 05$\h 35\m 14\fs21{-}14\fs64 and $\delta_{\mbox{\scriptsize J2000}} = -05\arcdeg 22\arcmin 32\arcsec{-}38\arcsec$ in the 0\farcs3 map) is 140 K, which agrees with methanol excitation temperatures toward the Compact Ridge reported by \citet{Crockett2014}. As seen in the full angular resolution \ce{^13CH3OH} excitation temperature map, the Hot Core-SW exhibits pockets of higher temperature, with some areas having temperatures as warm as $\sim$225 K. The sizes of these pockets range from $\sim$0\farcs15 to $\sim$0\farcs50 across, which corresponds to approximately 60 and 200 au, respectively, at Orion KL's distance. Thus, the pockets have the same spatial scales as embedded YSOs. 

Furthermore, Figure~\ref{fig:heatingsourceCH3OH} shows that the pockets of higher temperature align with regions where $\Delta T_{ex} < 0$. This suggests that there could be internal sources of heating at these sites, and that these warm pockets could house embedded YSOs which are heating the gas. 

However, these pockets are irregularly shaped and do not have temperature gradients resembling a localized ``spot heating'' effect, in which the gas temperature decreases radially outward from the center of an embedded YSO \citep[e.g.][]{vantHoff2018,vantHoff2020}. As such, we cannot definitively say whether these anomalous temperature pockets in the Hot Core-SW can be attributed to embedded sources of heating. Determining the nature of these pockets may require detailed radiative transfer modeling or additional continuum observations to complement the existing data sets. 

One of these pockets is co-spatial with a strong \ce{HCOOCH3} emission peak, reported as MF2 ($\alpha_{\mbox{\footnotesize J2000}} = 05$\h35\m14\fs44, $\delta_{\mbox{\footnotesize J2000}}= {-}05\arcdeg22\arcmin34\farcs4$) by \citet{Favre2011}. \citeauthor{Favre2011} report $T_{ex}\sim130$ K in both a 1\farcs8$\times$0\farcs8 ($T_{ex}=128\pm9$ K) and a 3\farcs6$\times$2\farcs2 ($140\pm14$ K) beam. As such, their results at this site do not agree with ours; however, their observations use larger beam sizes than our analysis, which likely contributes to this discrepancy. 

Like the Hot Core, the thermal structure of the Hot Core-SW region is likely affected by a mix of internal and external heating sources. At 0\farcs7 angular resolution, we see high temperatures of $\sim$300 K typical of hot cores adjacent to SMA1 and C22 and along the western edge of the Hot Core-SW. This may show the edge of this region being heated by the shocks from the Orion KL explosion propagating through the gas. This part of the Hot Core-SW is also marked by $\Delta T_{ex}>0$, further supporting some source of external heating. 

\subsection{IRc4}

IRc4 is a compact emission region north of the Compact Ridge that is associated with a high-density core on the order of $\sim$200-300 au \citep[][]{Hirota2015}. The heating source of IRc4 remains uncertain. \citet{Okumura2011} propose that IRc4 is a simple, externally heated dust cloud, whereas \citet{deBuizer2012} suggests that it is self-luminous. 

The \ce{^13CH3OH} $\Delta T_{ex}$ profile is ambiguous toward IRc4. East of the emission peak center, $\Delta T_{ex} < 0$ K, including when accounting for uncertainty, suggesting a source of internal heating. However, just west of center, $\Delta T_{ex} > 0$ within radii where simple Gaussian models of an internally heated source predict $\Delta T_{ex}<0$ K. These ambiguous results may indicate that IRc4 is indeed heated both internally close-in to the core and externally everywhere else. However, dedicated follow-up observations are necessary to confirm this hypothesis.

\subsection{Northwest Clump}

IRc6 is a mid-IR source toward the northwestern part of the Northwest Clump (NWC). In our observations, we do not detect vibrationally excited \ce{CH3CN}, which traces material pumped by far-IR radiation from massive stars, toward IRc6, and other studies have reported that there is no evidence of an embedded young star there \citep[e.g.][]{Shuping2004}.

Curiously, we find $\Delta T_{ex}<0$ K from the \ce{^13CH3OH} temperature measured toward IRc6. This signals that there is perhaps a source of internal heating in this region. 
Further evidence in support of IRc6 being internally heated can be seen in the \ce{^13CH3OH} excitation temperature maps, particularly the tapered angular resolution map (Figure~\ref{fig:resultslowresCH3OH}{)}. The Northwest Clump exhibits a radial temperature gradient from $\sim$95 K in the center to $\sim$65 K near the edges of that region. This localized ``spot heating'' effect, in which the gas temperature decreases radially outward from the center of an embedded young stellar object, has been observed toward the warm embedded disk in L1527 \citep{vantHoff2018} and the deeply embedded protostellar binary IRAS 16293-2422 \citep{vantHoff2020}. If there is, in fact, an internal heating source in the Northwest Clump, it likely does not play a dominant role in the thermal profile of the region \citep[][]{Li2020}. 



\section{Summary}\label{sec:summary}

We have used ground-state \ce{^13CH3OH} and vibrationally-excited \ce{CH3CN} emission in ALMA Band 4 from Cycles 4 and 5 observations to map the velocity and thermal structure of Orion KL at angular resolutions of ${\sim}0\farcs2{-}0\farcs3$ and ${\sim}0\farcs7{-}0\farcs9$. Overall, our results agree with the previous work at lower angular resolutions \citep[e.g., by][]{Wang2011,Friedel2011,Feng2015}. The angular resolution employed in this work probes spatial scales ${\le}350$ au in Orion KL, giving us a more localized look at the {nebula's} physical structure, {specifically that induced by YSOs.} 
\begin{enumerate}
\item We find evidence of possible sources of internal heating in the Hot Core. The velocity field exhibits an elongated structure centered on the Hot Core mm/submm emission peak. This elongated structure may be a low-velocity outflow with an LSR velocity of $\sim 7{-}8$ \kms, which is slightly higher than the ambient velocity ($V_{LSR}\sim5$ \kms) of the Hot Core region. Furthermore, the derived \ce{CH3CN} $\nu_8=1$ excitation temperature in the Hot Core is higher in a smaller beam than in a larger beam, suggesting an internal heating source. The thermal profile and apparent outflow structure emanating from the Hot Core may be evidence of an embedded protostar in the Hot Core.
\item The Hot Core-SW appears to have an exceptionally heterogeneous thermal structure. The average excitation temperature in the region is 140 K, and there are pockets of higher temperature as much as 225 K throughout. These pockets have higher temperatures in a smaller beam versus a larger beam, suggesting that they are internally heated. However, these pockets are not associated with compact mm/submm emission sources. The Hot Core-SW is also externally heated on the northeastern edge, presumably by the Orion KL explosion that took place about 500 years ago northwest of Source I. 
%
%
%
\item The thermal structure of IRc6 in the Northwest Clump suggests that this region has an internal heating source. Specifically, the temperature there is higher in a smaller beam, and the excitation temperature profile decreases radially from 95 K in the center to 65 K near the edges, resembling a spot heating effect found in embedded YSOs elsewhere. If there is an internal source of heating at IRc6, however, it does not play a dominant role in the Northwest Clump. 
\end{enumerate}

As we uncover more about Orion KL, it is becoming increasingly apparent that is a dynamic and heterogeneous region, and that each of its spatial components is not accurately described by one set of parameters. Furthermore, identifying the heating sources within Orion KL, and in particular the sources of internal heating, are critical to better understanding the nebula and its chemistry.

\acknowledgements{
This paper makes use of the following ALMA data: ADS/JAO.ALMA\#2017.1.01149, ADS/JAO.ALMA\#2016.1.01019, and ADS/JAO.ALMA\#2013.1.01034. 
ALMA is a partnership of ESO (representing its member states), NSF (USA), and NINS (Japan), together with NRC (Canada), MOST and ASIAA (Tawain), and KASI (Republic of Korea), in cooperation with the Republic of Chile. The Joint ALMA Observatory is operated by ESO, AUI/NRAO, and NAOJ. The National Radio Astronomy Observatory (NRAO) is a facility of the National Science Foundation (NSF) operated under Associated Universities, Inc. (AUI). This research made use of APLpy, an open-source plotting package for Python \citep{aplpy2012,aplpy2019}. 

This work has been supported by the NSF Graduate Research Fellowship Program under Grant No. DGE-1144469 and NRAO Student Observing Support under Award No. SOSPA6-014. OHW is additionally supported by an ARCS Los Angeles Founder Chapter scholarship. GAB gratefully acknowledges support from the NSF AAG (AST-1514918) and NASA Astrobiology (NNX15AT33A) and Exoplanet Research (XRP, NNX16AB48G) programs. This work benefited from discussions with Cam Buzard, Dana Anderson, Griffin Mead, Kyle Virgil, and Sadie Dutton. The authors thank Susanna Widicus Weaver and the anonymous referee for feedback on the prepared manuscript. OHW thanks Erica Keller, Sarah Wood, and the NRAO North American ALMA Science Center (NAASC) for their assistance with the data reduction. OHW also thanks the custodial staff at Caltech, especially Cruz Martinez. 
}

\facilities{ALMA}
\software{CASA \citep{McMullin2007}, CDMS \citep{Mueller2001}, LMFIT \citep[][]{lmfit}, APLpy \citep{aplpy2012,aplpy2019}, Gaussian16 \citep{g16}}

\appendix

\section{Assuming optically thin lines at local thermodynamic equilibrium}\label{sec:LTE}

The \ce{^13CH3OH} $\nu=0$ and \ce{CH3CN} $\nu_8=1$ transitions used in our analyses should be well-described by a local thermodynamic equilibrium (LTE). A rough collisional cross-section based on molecular geometry and van der Waals radii (2.096 \AA ~for \ce{^13CH3OH} and 2.408 \AA ~for \ce{CH3CN}, based on geometry optimization conducted through density functional theory (DFT) with the B3LYP level of theory and 6-31G basis set available in the \texttt{Gaussian} 16 computational chemistry software program) give critical densities on the order of $10^4{-}10^5$ cm$^{-3}$ for both compounds. The reported densities toward Orion KL are ${\ge}10^5$ cm$^{-3}$, suggesting that an LTE assumption is appropriate for these data. Moreover, LTE has been assumed in previous observations of complex organics toward Orion KL {due to the high density of Orion KL's substructures \citep[e.g.][]{Feng2015}}.

We assume optically thin lines at LTE following guidelines set forth by \citet{Goldsmith1999}. They write the optical depth $\tau$ of a transition as
\begin{equation}\label{eq:tauraw}
\tau = \frac{h}{\Delta V} N_u B_{ul}(e^{h\nu/kT}-1)
\end{equation}
where $\Delta V$ is the full-width at half-maximum line width in velocity units, $N_u$ is the column density in the upper state, $B_{ul}$ is the Einstein B coefficient for the transition, $\nu$ is the frequency of the transition, and $T$ is the excitation temperature for the molecule \citep[Eqn. 6]{Goldsmith1999}. This can be rewritten in terms of the Einstein A coefficient, which is related to $B_{ul}$ by
\begin{equation}\label{eq:einstein}
B_{ul} = \frac{A_{ul}c^3}{8\pi h\nu^3}.
\end{equation}
Moreover, $N_u$ can be written in terms of the total column density $N_{tot}$ as
\begin{equation}\label{eq:Nu}
N_u = \frac{N_{tot}}{Q(T)}g_u e^{-E_u/T}
\end{equation}
where $Q(T)$ is the excitation partition function, $g_u$ is the upper state degeneracy, and $E_u$ is the energy level of the upper state \citep[adapted from Eqn. 19]{Goldsmith1999}.
Substituting Eqns.~\ref{eq:einstein} and ~\ref{eq:Nu} into Eqn.~\ref{eq:tauraw}, the optical depth can be expressed as
\begin{equation}
\tau=\frac{g_u}{\Delta V}\frac{N_{tot}}{Q(T)}\frac{A_{ul}c^3}{8\pi\nu^3}e^{-E_u/T}(e^{h\nu/kT}-1).
\end{equation}

To test whether each transition of each molecular tracer is optically thin in our data, we calculated $\tau$ at each transition using the constants from Splatalogue listed in Table~\ref{tab:lines}. For \ce{^13CH3OH}, we selected a {spectrum extracted from a single beam area} near the center of the Compact Ridge, which has a total column density $N_{tot}$ of {$2.4\times10^{16}$ cm$^{-2}$}, a temperature of 196 K, and line width of 1.2 \kms; for \ce{CH3CN}, we selected a {spectrum extracted from a single beam area} near the center of the Hot Core, where $\Delta V=2.9$ \kms, {$N_{tot}=3.9\times10^{16}$ cm$^{-2}$}, and $T_{ex}=275 K$. All lines have $\tau\le 0.02\ll1$ thus are optically thin.

Furthermore, we ran $\chi^2$ tests as an additional check for whether the LTE model is appropriate for the data presented in this work. The average $\chi^2$ values for the \ce{^13CH3OH} fits are 1.02 and 1.85 for the 0\farcs3 and 0\farcs7 data, respectively. For \ce{CH3CN} $\nu_8=1$, $\chi^2$ is 3.60 and 0.15 for the 0\farcs2 and 0\farcs9 maps, respectively. These are slightly better than the $\chi^2$ calculations for these molecules by \citet{Crockett2014}, who calculated 1.9 for \ce{^13CH3OH} and 4.0 for \ce{CH3CN} $\nu_8=1$. 

\section{Uncertainty maps}\label{sec:uncertainty}
\restartappendixnumbering

\begin{figure}
    \centering
    \epsscale{1.5}
    \plottwo{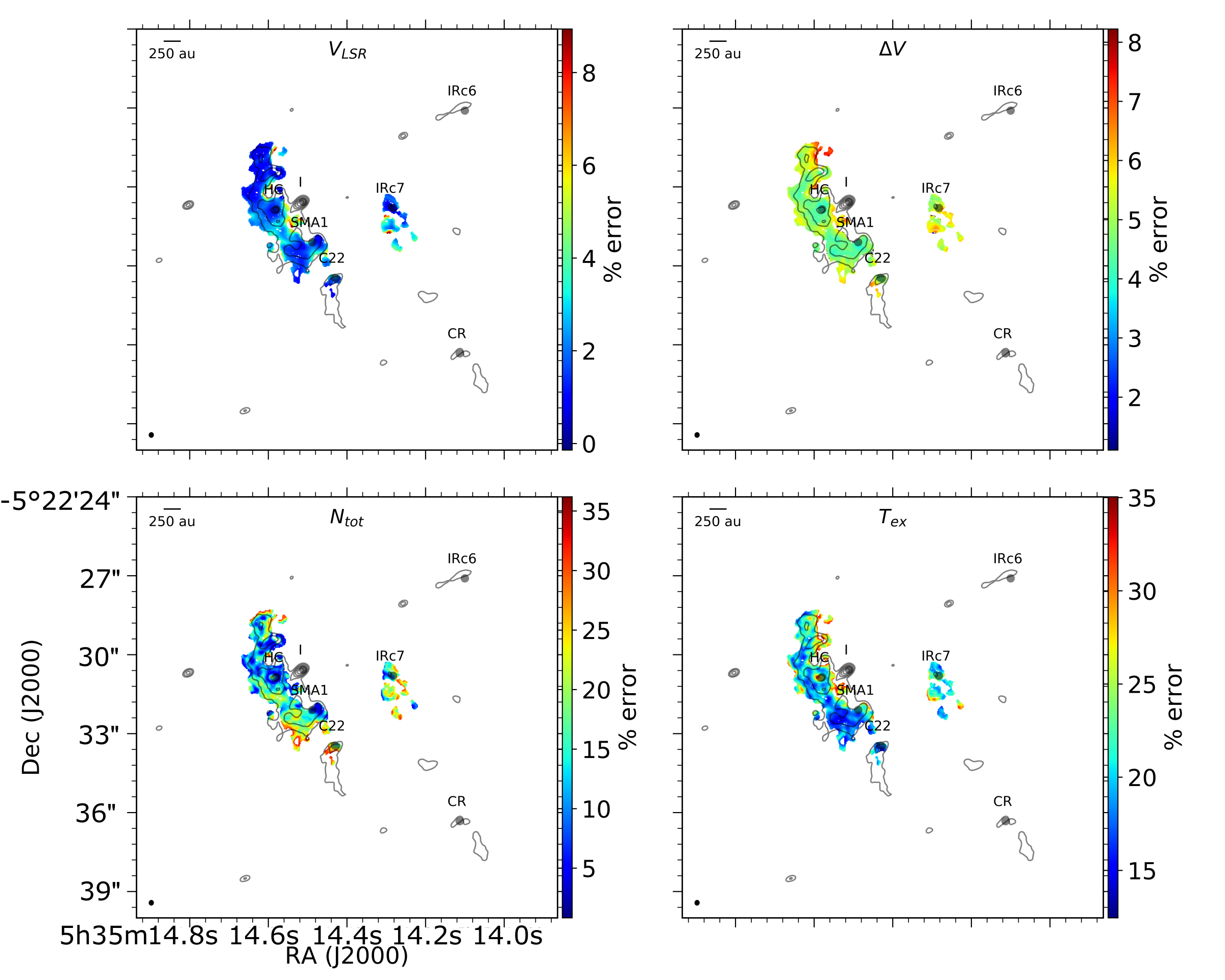}{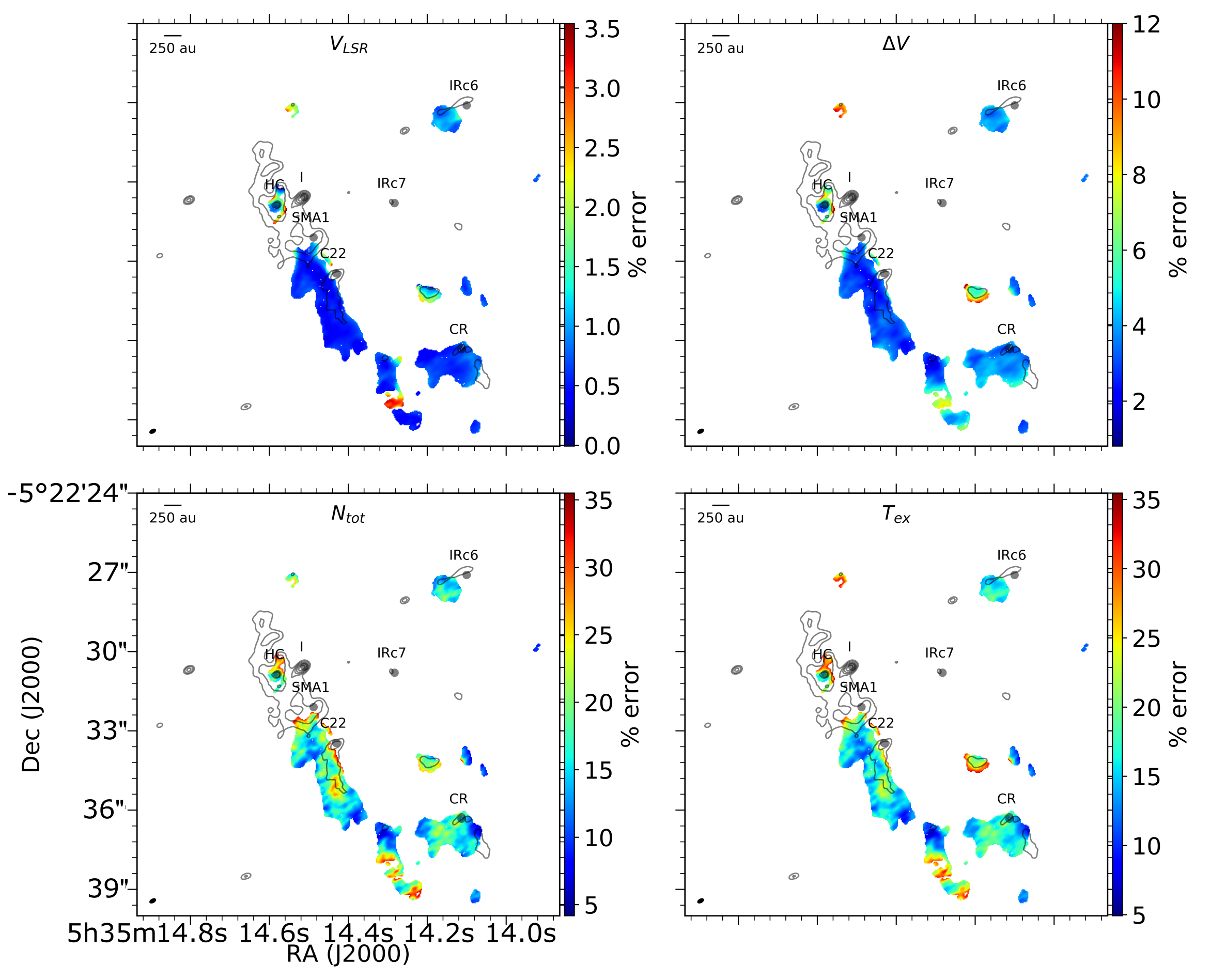}
    \caption{{Propagated percent error uncertainty maps for Figures~\ref{fig:resultshiresCH3CN} and~\ref{fig:resultshiresCH3OH}, showing uncertainties for full resolution (first row) \ce{CH3CN} $\nu_8 = 1$ velocity field and line width field; (second row) \ce{CH3CN} $\nu_8=1$ total column density and excitation temperature; (third row) \ce{^13CH3OH} velocity field and line width field; and (fourth row) \ce{^13CH3OH} total column density and excitation temperature. Contours show the 2 mm ($\sim$150 GHz) continuum from the left panel of Figure~\ref{fig:continuum}.}}
    \label{fig:pcterrhigh}
\end{figure}

\begin{figure}
    \centering
    \epsscale{1.5}
    \plottwo{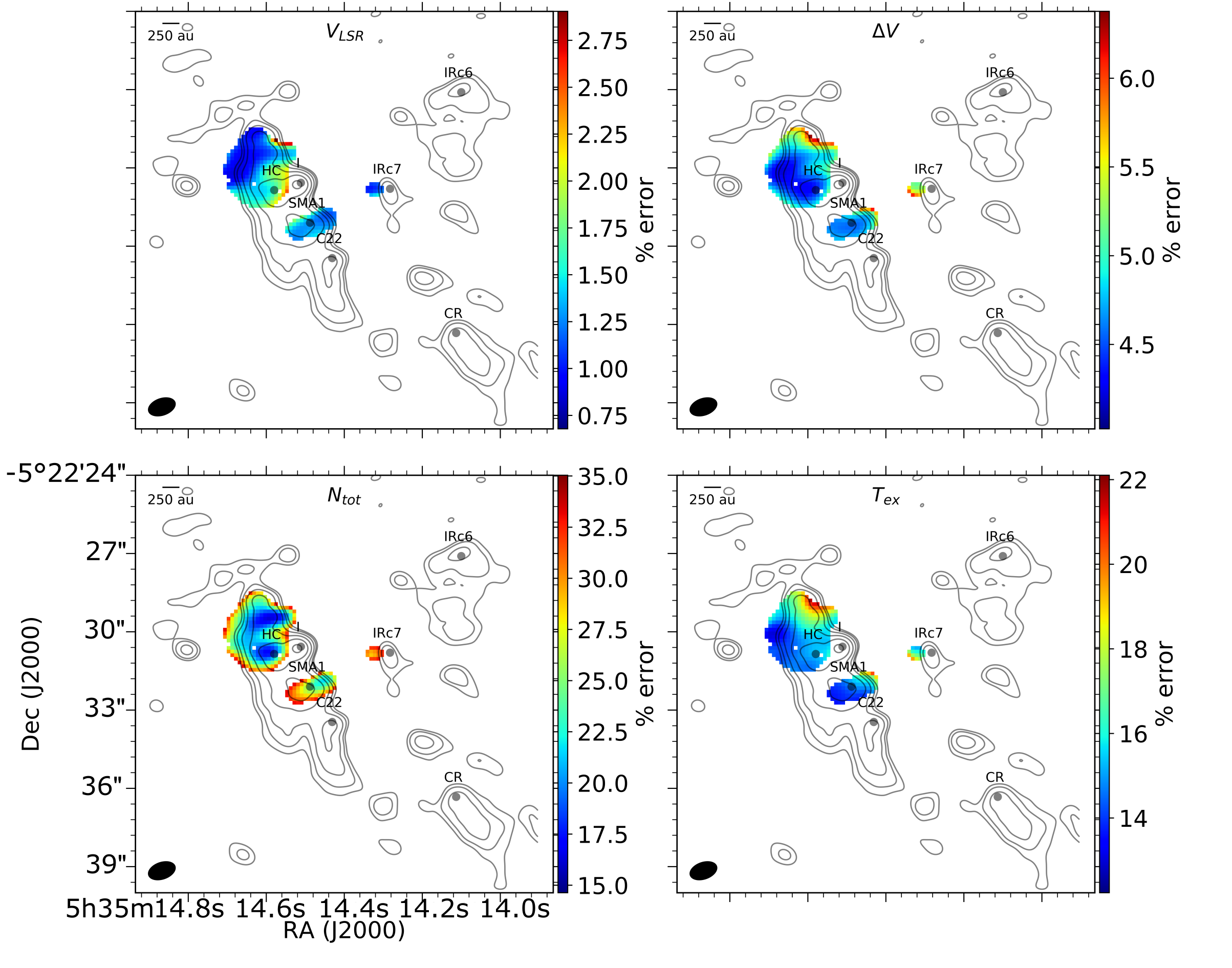}{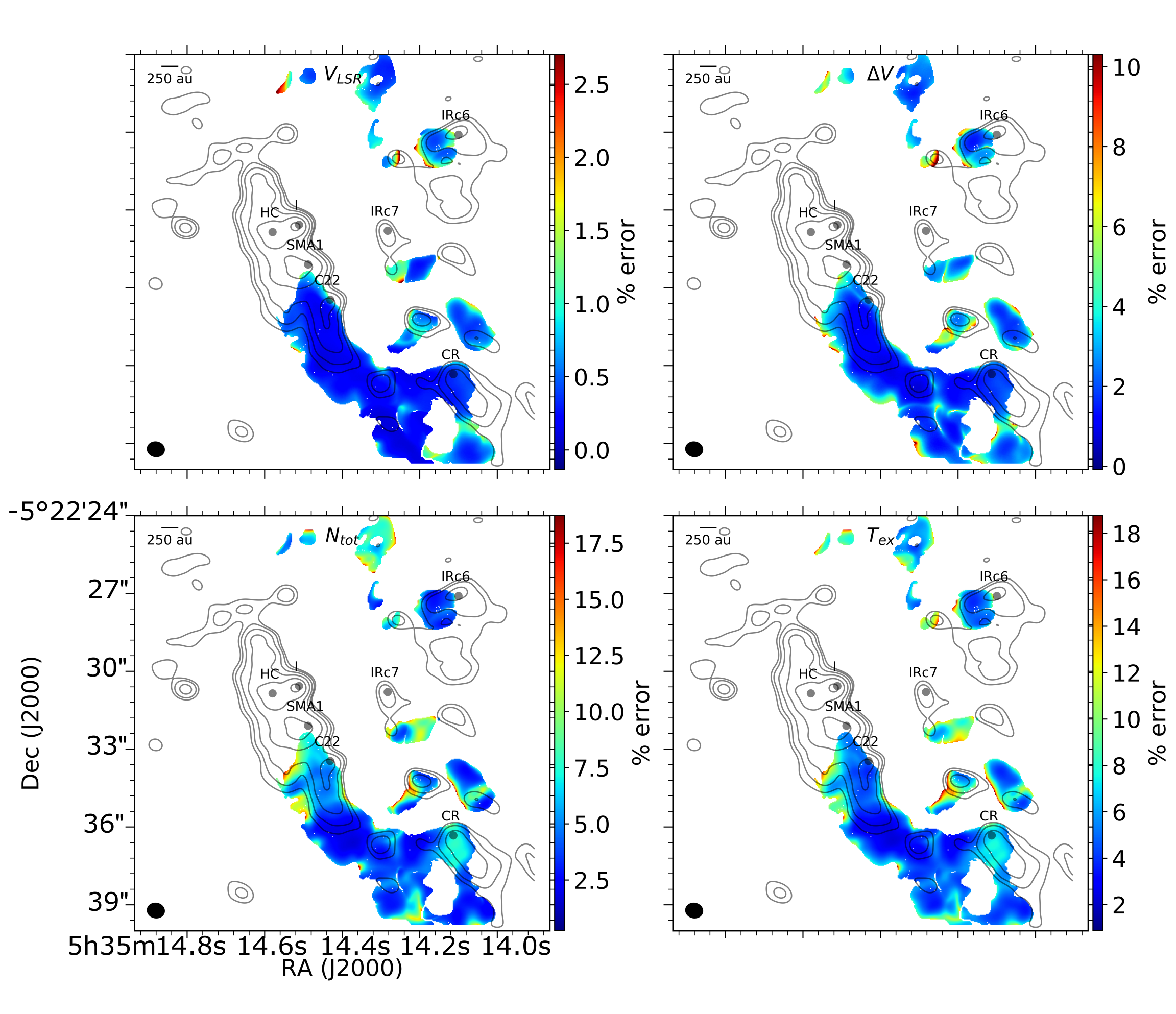}
    \caption{{Propagated percent error uncertainty maps for Figures~\ref{fig:resultslowresCH3CN} and ~\ref{fig:resultslowresCH3OH}, showing uncertainties for tapered resolution (first row) \ce{CH3CN} $\nu_8 = 1$ velocity field and line width field; (second row) \ce{CH3CN} $\nu_8=1$ total column density and excitation temperature; (third row) \ce{^13CH3OH} velocity field and line width field; and (fourth row) \ce{^13CH3OH} total column density and excitation temperature. Contours show the 2 mm ($\sim$150 GHz) continuum from the right panel of Figures~\ref{fig:continuum}.}}
    \label{fig:pcterrlow}
\end{figure}

Percent uncertainty maps were calculated from the derived \texttt{LMFIT} standard errors. These maps are presented in Figures~\ref{fig:pcterrhigh} and~\ref{fig:pcterrlow} for the full and tapered angular resolution images, respectively.

\section{Internal versus external heating profile models}\label{sec:toymodels}

\begin{figure}
    \figurenum{C1}
    \centering
    \gridline{\fig{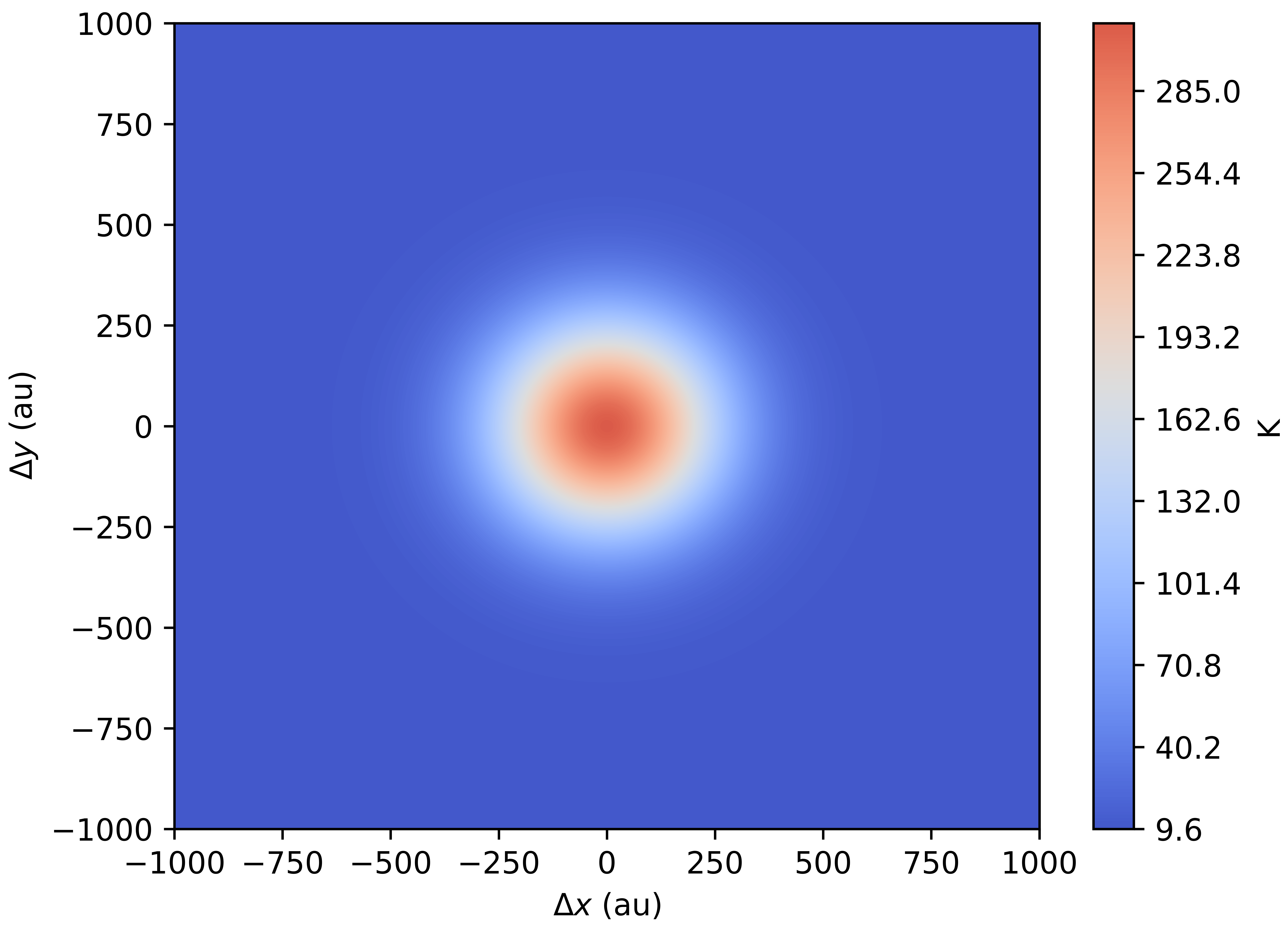}{0.5\textwidth}{(a)}
    \fig{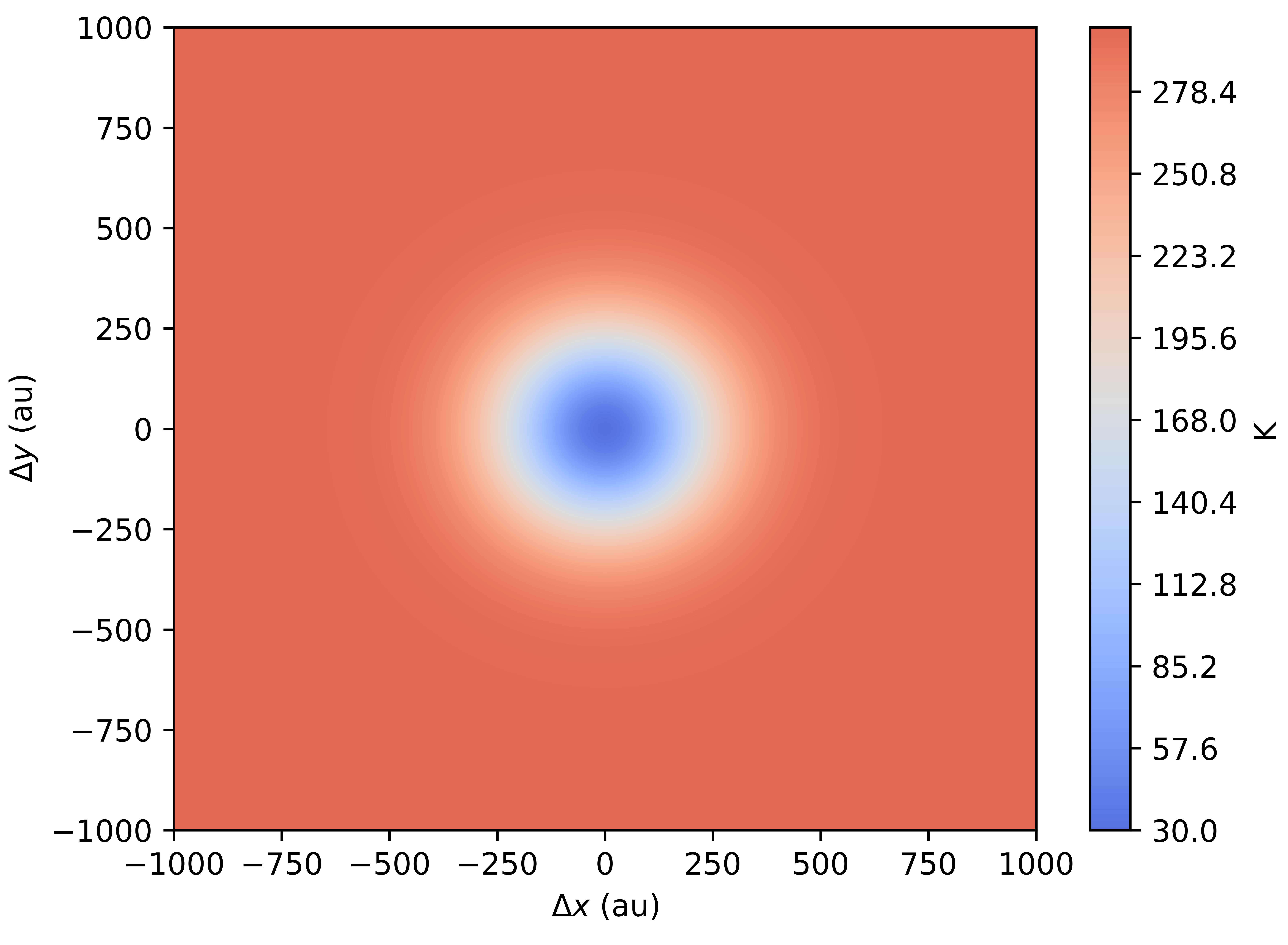}{0.5\textwidth}{(b)}}
    \gridline{\fig{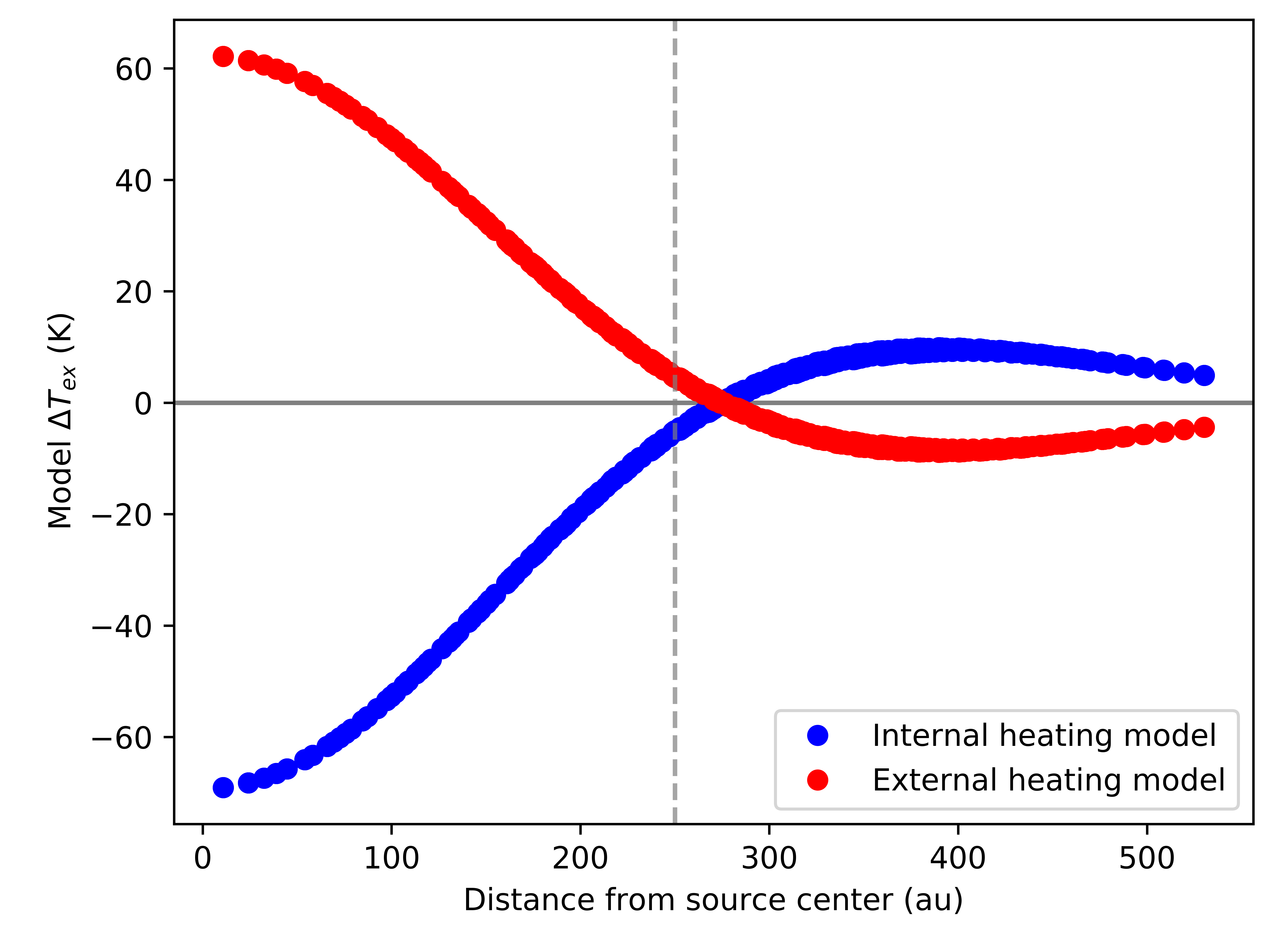}{0.5\textwidth}{(c)}}
    \caption{{Example of the simple models used to test the sign of $\Delta T_{ex}$. These plots show the temperature $T_{ex}$ profile for an internally (a) and an externally (b) heated source with a radius of 250 au. The value of $\Delta T_{ex}$ (c) for each profile was measured at different radii from the source center. Here, the plot shows $\Delta T_{ex}$ with a small beam of 0\farcs2 and a large beam of 0\farcs9 (same as that used in the \ce{CH3CN} $\nu_8=1$ images.)}}
    \label{fig:toymodel}
\end{figure}

We ran simple models for different source sizes (with radii between 100 and 1000 au) and temperature gradients to test whether internally and externally heated sources are consistently characterized by $\Delta T_{ex}<0$ and $\Delta T_{ex}>0$, respectively. The models calculated the temperature inside a full and tapered synthesized beam at various radii across 2-D Gaussian temperature profiles (both circular and elliptical). The models were run assuming both a smooth (unchanging) temperature background and a linear temperature gradient background. In all models, the sign of $\Delta T_{ex}$ accurately reflected whether a source of radius $r_{src}$ was internally or externally heated at radii within $0.67r_{src}$ of the source center, with `higher contrast' source profiles (i.e., sources that have either much higher peak temperatures than their surroundings or particularly steep temperature gradients) being more likely to deviate from the predicted sign of $\Delta T_{ex}$ toward their outer radii. Smaller sources (e.g., with radii of 100 au) consistently had $\Delta T_{ex}$ values in agreement with their source of heating regardless of temperature contrast. In other words, within 100 au or more of a source's center (e.g., mm emission peak), the $\Delta T_{ex}$ sign is expected to accurately reflect whether that source is internally or externally heated. {An example of the simple models used to test the sign of $\Delta T_{ex}$ is shown in Fig.~\ref{fig:toymodel}.}

\newpage
\bibliography{orionbib}{}
\bibliographystyle{aasjournal}

\listofchanges
\end{document}